\documentclass{article}
\usepackage{amsthm,amsmath}
\usepackage[utf8]{inputenc} 
\usepackage{amsfonts}
\usepackage{algorithm}
\usepackage{algorithmic}
\usepackage{multirow}
\usepackage{url}
\usepackage{authblk}
\usepackage{graphicx}
\usepackage{tabularx}
\usepackage{color}
\newcommand{\reals}{\mathbb{R}}
\newcommand{\softmax}{\ensuremath{\mathrm{softmax}}}
\DeclareMathOperator*{\argmax}{arg\,max}
\newtheorem{observation}{Observation}[section]

\newcommand{\todo}[1]{\textcolor{red}{*TODO: #1*}}
\begin{document}
\title{Complex Network Effects on the Robustness of Graph Convolutional Networks\thanks{This material is based upon work supported by the United States Air Force under Air Force Contract No. FA8702-15-D-0001 and the Combat Capabilities Development Command Army Research Laboratory (under Cooperative Agreement Number W911NF-13-2-0045). Any opinions, findings, conclusions or recommendations expressed in this material are those of the authors and do not necessarily reflect the views of the United States Air Force or Army Research Laboratory.}}
\date{}

\author[1]{Benjamin~A.~Miller\thanks{Contact author: \texttt{miller.be@northeastern.edu}}}
\author[2]{Kevin~Chan}
\author[1]{Tina~Eliassi-Rad}
\affil[1]{Northeastern University, Boston MA}
\affil[2]{Army Research Laboratory, Adelphi MD}
\maketitle
\begin{abstract}
        Vertex classification---the problem of identifying the class labels of nodes in a graph---has applicability in a wide variety of domains. Examples include classifying subject areas of papers in citation networks or roles of machines in a computer network. Vertex classification using graph convolutional networks is susceptible to targeted poisoning attacks, in which both graph structure and node attributes can be changed in an attempt to misclassify a target node. This vulnerability decreases users’ confidence in the learning method and can prevent adoption in high-stakes contexts. Defenses have also been proposed, focused on filtering edges before creating the model or aggregating information from neighbors more robustly. This paper considers an alternative: we leverage network characteristics in the training data selection process to improve robustness of vertex classifiers.
    
    We propose two alternative methods of selecting training data: (1) to select the highest-degree nodes and (2) to iteratively select the node with the most neighbors minimally connected to the training set. In the datasets on which the original attack was demonstrated, we show that changing the training set can make the network much harder to attack. To maintain a given probability of attack success, the adversary must use far more perturbations; often a factor of 2--4 over the random training baseline. In addition, these training set selection methods often work in conjunction with the best recently published defenses to provide even greater robustness. While increasing the amount of randomly selected training data sometimes results in a more robust classifier, the proposed methods increase robustness substantially more. We also run a simulation study in which we demonstrate conditions under which each of the two methods outperforms the other, controlling for the graph topology, homophily of the labels, and node attributes.
\end{abstract}
\section{Introduction}
\label{sec:intro}
Classification of vertices in graphs is an important problem in a variety of applications, from e-commerce (classifying users for targeted advertising) to security (classifying computer nodes as malicious or not) to bioinformatics (classifying roles in a protein interaction network). In the past several years, numerous methods have been developed for this task (see, e.g.,~\cite{graphsage, Moore2017}). More recently, research has focused on attacks by adversaries~\cite{Zugner2018, Dai2018} and robustness to such attacks~\cite{Wu2019}. If an adversary were able to insert misleading data into the training set (e.g., generate benign traffic during a data collection period that could conceal its behavior during testing/inference time), the chance of successfully evading detection would increase, leaving data analysts unable to respond to potential threats. 

To classify vertices in the presence of adversarial activity, we must implement learning systems that are robust to such potential manipulation. If such malicious behavior has low cost to the attacker and imposes high cost on the data analyst, machine learning systems will not be trusted and adopted for use in practice, especially in high-stakes scenarios such as network security and traffic safety. Understanding how to achieve robustness is key to realizing the full potential of machine learning.

Adversaries, of course, will attempt to conceal their manipulation. The first published poisoning attack against vertex classification was an adversarial technique called Nettack~\cite{Zugner2018}, which can create perturbations that are subtle while still being extremely effective in decreasing performance on the target vertices. The authors use their poisoning attack against a graph convolutional network (GCN). 

From a defender's perspective, we aim to make it more difficult for the attacker to cause node misclassification. In addition to changing the properties of the classifier itself, there may be portions of a complex network that provide more information for learning than others. Complex networks are highly heterogeneous and random sampling may not be the best way to obtain labels. If there is flexibility in the means of obtaining training data, the defender should leverage what is known about the graph topology.

This paper demonstrates that leveraging complex network properties can improve robustness of GCNs in the presence of adversaries. We focus on two alternative techniques for training data selection. Both methods aim to train with a subset of nodes that are well connected to the held out set. Here we see a benefit, often raising the number of perturbations required for a given level of attack success by over a factor of 2. When it is possible to pick a specific subset on which to train, this can provide a significant advantage. Some combination of these methods will likely be useful to develop a more robust vertex classification system.

\subsection{Scope and Contributions}
In this paper, we are specifically interested in \emph{targeted poisoning} attacks against vertex classifiers, where the data are modified at training time to cause a specific target node to be misclassified. We consider attacks against the structure of the graph, rather than against node attributes. We focus on classification methods where there is an implicit assumption of homophily. Working within this context, the contributions of this work are as follows:
\begin{itemize}
    \item We propose two methods---\textsc{StratDegree} and \textsc{GreedyCover}---for selecting training data that result in a greater burden on attackers.
    \item We demonstrate that the robustness gained via these methods cannot be reliably obtained by simply increasing the amount of randomly selected training data.
    \item We show that the most robust defense methods are often improved by working in conjunction with \textsc{GreedyCover}.
    \item We show that there is no consistent tradeoff between the robustness gained from these methods and classification performance.
    \item In simulation, we study the effects of various generative models and report the impact of class homophily, topological features, and node attribute similarity across classes on classification performance and robustness to attack.
    
\end{itemize}
These contributions all point toward interesting future research in this area, such as determining the conditions under which such methods are effective.

\subsection{Paper Organization}
The remainder of this paper is organized as follows. In Section~\ref{sec:related}, we briefly contextualize our work within the current literature. In Section~\ref{sec:model} we describe the vertex classification problem, GCNs, and the Nettack method. Section~\ref{sec:methods} outlines the methods we investigate to select training data, and Section~\ref{sec:setup} details the experimental setup, including datasets, attacks, and classification methods. Section~\ref{sec:results} documents experimental results on real data, illustrating the effectiveness of the proposed methods. In Section~\ref{sec:simulation}, we present the results of a simulation study in which we vary graph topology, node attributes, and homophily, and evaluate robustness of the methods across the landscape. In Section~\ref{sec:conclusion} we conclude with a summary and outline open problems and future work.

\section{Related Work}
\label{sec:related}
Adversarial examples in deep neural networks have received considerable attention since they were documented several years ago~\cite{Szegedy2014}. Since that time, numerous attack methods have been proposed, largely focused on the image classification domain (though there has been interest in natural language processing as well, e.g.,~\cite{Jia2017}). In addition to documenting adversarial examples, Szegedy et al. demonstrated that such examples can be generated using the limited-memory BFGS (L-BFGS) algorithm, which identifies an adversarial example in an incorrect class with minimal $L_2$ norm distance to the true data. Later, Goodfellow et al. proposed the fast gradient sign  method (FGSM), where the attacker starts with a clean image and takes small, equal-sized steps in each dimension (i.e., alters each pixel by the same amount) in the direction maximizing the loss~\cite{Goodfellow2015}. Another proposed attack---the Jacobian-based Saliency Map Attack (JSMA)---iteratively modifies the pixel with the largest impact on the loss~\cite{Papernot2016a}. DeepFool, like L-BFGS, minimizes the $L_2$ distance from the true instance while crossing a boundary into an incorrect class, but does so quickly by approximating the classifier as linear, stepping to maximize the loss, then correcting for the true classification surface~\cite{Moosavi-Dezfooli2016}. Like Nettack, these methods all try to maintain closeness to the original data ($L_2$ norm for L-BFGS and DeepFool, $L_0$ norm for JSMA, and $L_\infty$ norm for FGSM).

Some of these methods have been adapted for use with graph data. 
In~\cite{WuH2019}, the authors modify FGSM and JSMA to use integrated gradients and show it to be effective against vertex classification. 
In addition, new attacks against vertex classification have been introduced, including a method that uses reinforcement learning to identify modifications to graph structure for an evasion attack~\cite{Dai2018}. 
To increase the scale of attacks, the authors of~\cite{Li2021} propose an attack that only considers a $k$-hop neighborhood of the target. This method attacks a simplified GCN, introduced in~\cite{WuF2019}, which applies a logistic regression classifier after $k$ rounds of feature propagation.

Defenses to attacks such as FGSM and JSMA have been proposed, although several prove to be insufficient against stronger attacks. A simple improvement is to include adversarial examples in the training data~\cite{Goodfellow2015}. Defensive distillation is one such defense, in which a classifier is trained with high ``temperature'' in the softmax, which is reduced for classification~\cite{Papernot2016}. While this was effective against the methods from~\cite{Szegedy2014,Goodfellow2015, Papernot2016a,Moosavi-Dezfooli2016}, it was shown in~\cite{Carlini2017} that modifying the attack by changing the constraint function (which ensures the adversarial example is in a given class) renders this defense ineffective. More defenses have been proposed, such as pixel deflection~\cite{Prakash2018} and randomization techniques~\cite{Xie2018}, but many such methods are still found to be vulnerable to attacks~\cite{Athalye2018,Athalye2018a}. Other work has focused on provably robust defenses~\cite{Wong2018}, with empirical performance often close to certifiable claims~\cite{Croce2019}. Stochastic networks have also shown improved robustness to various attacks~\cite{Dapello2021}. In the wake of growing interest in adversarial robustness, several authors in the community have aggregated best practices for evaluation of systems in~\cite{Carlini2019}.

More recent work has focused on robustness of GCNs, including work on robustness to attacks on attributes~\cite{Zugner2019b} and more robust GCN variants~\cite{Zhu2019}. Multiple authors have considered aggregation techniques that are less sensitive to outliers~\cite{Geisler2020, Chen2021}. One approach to a more robust classifier incorporates an attention mechanism that learns the importance of the attributes of other nodes' features to a node's class~\cite{Velivckovic2018}. Others have considered using a GCN with modified graph structure to improve robustness, such as using a low-rank approximation for the graph~\cite{Entezari2020} and filtering edges based on attribute values~\cite{WuH2019}. Another method also considers attribute values, in this case creating a similarity graph from the attributes that augments the given graph structure to preserve node similarity in feature space~\cite{Jin2021}. The low-rank structure and node similarity concepts are combined in~\cite{Jin2020a} to create a neural network that aims to simultaneously learn the true graph structure from poisoned data and learn a classifier of unlabeled nodes. The authors of~\cite{Dai2022} explore a similar idea in the context of noisy data and few labels, using link prediction to augment the observed graph. Relations between attribute similarity and node class---including possible heterophily---are also exploited in \textsc{GNNGuard}~\cite{Zhang2020}. More recent work has shown that attacks that are adaptive to defenses easily undermine the robustness increase observed when using non-adaptive attacks~\cite{Mujkanovic2022}. 
Other recent GCN developments include modifications to deal with heterophily, via classifier design choices~\cite{Zhu2020} and by learning the level of homophily or heterophily in the graph as part of the training procedure~\cite{Zhu2021}. Several attacks~\cite{Dai2018,Zugner2018,Chen2018,Zugner2019a,WuH2019,Xu2019} and defenses~\cite{Goodfellow2015,WuH2019,Zhu2019,Entezari2020,Jin2020a} have been incorporated into a software package called DeepRobust, enabling convenient experimentation across a variety of conditions~\cite{LiY2020, Jin2020b}. As with neural networks more generally, there has been work on certifiable robustness for GCNs~\cite{Bojchevski2019b, Zugner2020}.

While this paper is focused on targeted attacks, several attacks, such as~\cite{Zugner2019a, Xu2019}, attack the whole graph in order to degrade overall performance. Some attacks in this area have allowed adding new nodes~\cite{Sun2019}, flipping labels~\cite{Liu2019}, and rewiring edges~\cite{Ma2019}. In addition, there are many machine learning tasks on graphs other than vertex classification, and work has been done on, for example, edge classification in an adversarial context~\cite{Yu2018}.
\section{Problem Model}
\label{sec:model}
We consider the vertex classification problem as described in~\cite{Zugner2018}, where we are given an undirected graph $G=(V,E)$ of size $N=|V|$ and an $N\times d$ matrix of vertex attributes $X$. Each node has an arbitrary numeric index from 1 to $N$. For this work, as in~\cite{Zugner2018}, we consider only binary attributes. In addition to its $d$ attributes, each node has a label denoting its class. We enumerate classes as integers from $1$ to $C$. Given a subset of labeled instances, the goal is to correctly classify the unlabeled nodes.

The focus of~\cite{Zugner2018} is on GCNs, which make use of the adjacency matrix for the graph $A=\{a_{ij}\}$, where $a_{ij}$ is 1 if there is an edge between node $i$ and node $j$ and is 0 otherwise. The GCN applies a symmetrized one-hop graph convolution~\cite{Kipf2017} to the input layer. That is, if we let $D$ be the diagonal matrix of vertex degrees---i.e., the $i$th diagonal entry is the number of edges connected to vertex $i$, $d_{ii}=\sum_{j=1}^N{a_{ij}}$---then the output of the first layer of the network is expressed as
\begin{equation}
    H = \sigma\left(D^{-1/2}AD^{-1/2}XW_1\right),\label{eq:graphConv}
\end{equation}
where $W_1$ is a weight matrix, $X$ is a feature matrix whose $i$th row is $x_i^T$ (the attribute vector for row vertex $i$), and $\sigma$ is the rectifier function. From the hidden layer to the output layer, a similar graph convolution is performed, followed by a softmax output:
\begin{equation*}
    Y = \softmax\left(D^{-1/2}AD^{-1/2}HW_2\right).\
\end{equation*}
The focus in~\cite{Zugner2018} is on GCNs with a single hidden layer. Each vertex is then classified according to the largest entry in the corresponding row of $Y$.

The vertex attack proposed in~\cite{Zugner2018} operates on a surrogate model where the rectifier function is replaced by a linear function, thus approximating the overall network as
\begin{align}
    Y&\approx \softmax\left(\left(D^{-1/2}AD^{-1/2}\right)^2XW_1W_2\right)\label{eq:surrogate}\\
    &=\softmax\left(\left(D^{-1/2}AD^{-1/2}\right)^2XW\right).\nonumber
\end{align}
Nettack uses a greedy algorithm to determine how to perturb both $A$ and $X$ to make the GCN misclassify a target node. The changes are intended to be ``unnoticeable,'' i.e., the degree distribution of $G$ and the co-occurrence of features are changed negligibly. Using the approximation in (\ref{eq:surrogate}), Nettack perturbs by either adding or removing edges or turning off binary features so that the classification margin is reduced the most at each step. Note that while it can change the topology and the features, Nettack does \emph{not} change the labels of any vertices. In this paper, we only consider structural perturbations. Nettack allows either \emph{direct} attacks, in which the target node itself has its edges and features changed, or indirect \emph{influence} attacks, where neighbors of the target have their data altered.

The classifier is evaluated in a context where only some of the labels are known, and the labeled data are split into training and validation sets. 
To train the GCN, 10\% of the data are selected at random (or by one of the alternative methods outlined in Section~\ref{sec:methods}), and another 10\% is selected for validation. The remaining 80\% is the test data. After training, nodes are selected for attack among those that are correctly classified. The goal of the defender is to make the a successful attack as expensive as possible.

As we discuss in Section~\ref{sec:setup}, we also consider attacks other than Nettack, and classifiers other than standard GCNs. While the details differ (e.g., using different criteria to identify perturbations), the overall problem model remains the same.

\section{Proposed Methods}
\label{sec:methods}

As we investigated classification performance using Nettack, we noted that nodes in the test set with many neighbors in the training set were more likely to be correctly classified. This dependence on labeled neighbors is consistent with previous observations~\cite{Neville2009}. We observed this effect using the standard method of training data selection used in the original Nettack paper: randomly select 10\% for training, 10\% for validation, and 80\% for testing. This observation suggested that a training set where the held-out nodes are well represented among neighborhoods of the training data---providing a kind of ``scaffolding'' for the unlabeled data---could make the classification more robust.


We considered two methods to test this hypothesis. The first simply chooses the highest-degree nodes (stratified by class) to be in the training set.  We refer to the stratified degree-based thresholding method as \textsc{StratDegree}. The other method uses a greedy approach in an attempt to ensure every node has at least a minimal number of neighbors in the training set. Starting with an empty training set and a threshold $k=0$, we iteratively add a node of a particular class with the largest number of neighbors that are connected to at most $k$ nodes in the training set. The class is randomly selected based on how many nodes of each class are currently in the training set and the number required to achieve class stratification. When there are no such neighbors, we increment $k$. This procedure continues until we have the desired proportion of the overall dataset for training. Algorithm~\ref{alg:cover} provides the pseudo-code. 

\begin{algorithm}
\begin{centering}
\begin{algorithmic}
\STATE \textbf{Input:} Graph $G=(V,E)$, training proportion $t\in (0,1)$, classes $C$, class map $\textrm{class}:V\rightarrow C$
\STATE \textbf{Output:} Training set $T\subset V$
\STATE $k\gets 0$
\FORALL{$u\in V$}
\STATE $m_u\gets 0$ \ \ \ \ $\langle\langle$number of neighbors in the training set: initialize all nodes to 0$\rangle\rangle$
\ENDFOR
\FORALL{$c\in C$}
\STATE $\langle\langle$initialize vertex and training subsets partitioned by class$\rangle\rangle$
\STATE $V_c\gets \{v\in V | \textrm{class}(v)=c\}$
\STATE $T_c\gets\emptyset$
\ENDFOR
\WHILE{$|T_c|<t|V_c|~\forall c\in C$}
    \STATE $c\gets$ class selected with probability proportional to $1-\frac{|T_c|}{t|V_c|}$
    \STATE $v \gets \argmax_{u\in V_c\setminus T_c} \sum_{u^\prime\in\mathcal{N}(u)}\mathbb{I}\left[m_{u^\prime}=k\right]$
    \IF{$\sum_{u^\prime\in\mathcal{N}(v)}\mathbb{I}\left[m_{u^\prime}=k\right]=0$}
        \STATE {$k\gets k+1$} \ \ $\langle\langle$incr. min. num. trained neighbors$\rangle\rangle$
    \ELSE
        \STATE{$T_c\gets T_c\cup \{v\}$}
        \STATE$T\gets\bigcup_{c\in C}{T_c}$
        \STATE $m_v\gets -1$
        \FORALL {$u^\prime\in \mathcal{N}(v)\setminus T$}
            \STATE $m_{u^\prime}\gets m_{u^\prime}+1$
        \ENDFOR
    \ENDIF
\ENDWHILE
\RETURN $T$
\end{algorithmic}
\caption{\textsc{GreedyCover}}
\label{alg:cover}
\end{centering}
\end{algorithm}


\subsection{Computational Complexity}
Using \textsc{StratDegree} and \textsc{GreedyCover} both have computational costs beyond random sampling. \textsc{StratDegree} requires finding the highest-degree nodes, which, for a constant fraction of the dataset size, will require $O(|E|+|V|\log{|V|})$ time (for computing degrees and sorting), compared to $O(|V|)$ time for random sampling. Each step in \textsc{GreedyCover} requires finding the vertex with the most neighbors minimally connected to the training set. As written in Algorithm~\ref{alg:cover}, each iteration requires $O(|E|)$ time to count the number of such neighbors each node has, which would result in an overall running time of $O(|V||E|)$. This could be improved using a priority queue---such as a Fibonacci heap---to achieve $O(|E|+|V|\log{|V|})$ time ($O(|V|)$ logarithmic-time extractions of the minimum and $O(|E|)$ constant-time key updates. Thus, the two proposed method require moderate overhead compared to the running time for the GCN.
\section{Experimental Setup}
\label{sec:setup}

\begin{figure}
    \centering
    \includegraphics[width=0.95\textwidth]{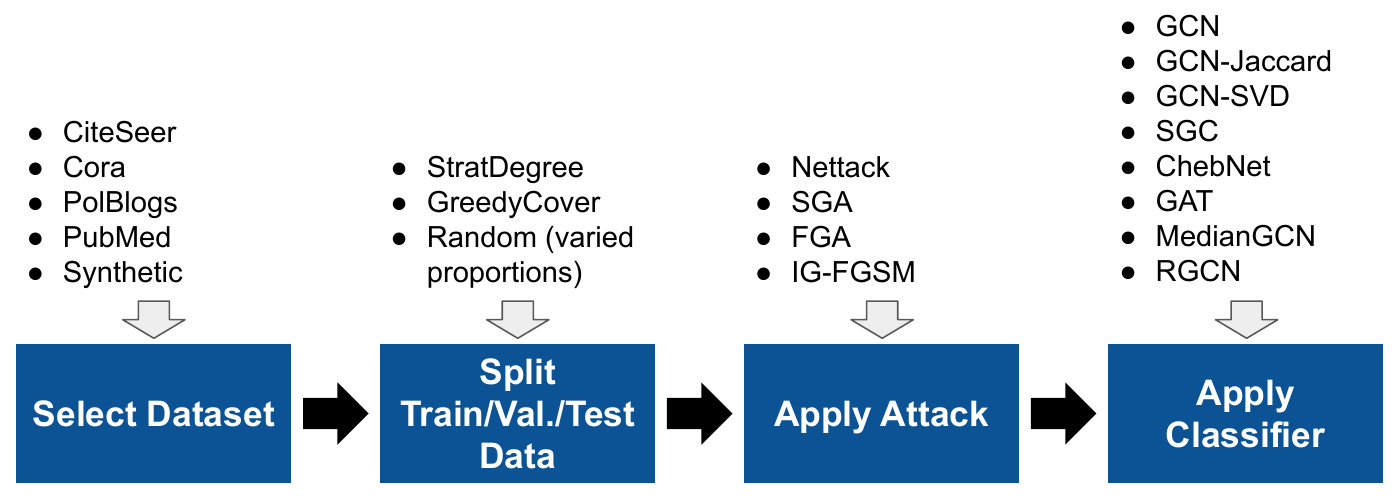}
    \caption{Processing chain for experiments. Each experiment takes a dataset, applies some method to split training, validation, and test data, applies an attack to a set of target nodes, then applies a classifier to the attacked dataset. We evaluate the robustness of vertex classification----in terms of required attacker budget at a given attack success rate---across all possible combinations of dataset, selection methods, attacks, and classifier.}
    \label{fig:procChain}
\end{figure}

Each experiment in our study involves (1) a graph dataset, (2) a method for selecting training data, (3) a structure-based attack against vertex classification, and (4) a classification algorithm. We consider several options for each step in this process, as shown in Figure~\ref{fig:procChain}. This section details the methods and datasets we use across the experiments in this paper. We use the DeepRobust library~\cite{LiY2020} for datasets, attacks, and classifiers.

\subsection{Datasets}
We use the three datasets used in the Nettack paper in our experiments, plus one larger citation dataset:
\begin{itemize}
    \item \textbf{CiteSeer} The CiteSeer dataset has 3312 scientific publications put into 6 classes. The network has 4732 links representing citations between the publications. The features of the nodes contain 1s and 0s indicating the presence of the word in the paper. There are 3703 unique words considered for the dictionary.
    
    \item \textbf{Cora} The Cora dataset consists of 2708 machine learning papers classified into one of seven categories. The citation network consists of 5429 citations. For each paper (vertex) in the network there is a feature vector of 0s and 1s for whether it contains one of 1433 unique words.
    
    \item \textbf{PolBlogs} The political blogs dataset consists of 1490 blogs labeled as either liberal or conservative. A total of 19,025 links between blogs form the directed edges of the graph. No attributes are used.
    
    \item \textbf{PubMed} The PubMed dataset consists of 19,717 papers pertaining to diabetes classified into one of three classes. The citation network consists of 44,338 citations. For each paper in the network there is a binary feature vector representing the presence of 500 words.
\end{itemize}

\subsection{Training Data Selection}
\label{subsec:dataselection}
To select training data, we use \textsc{StratDegree} and \textsc{GreedyCover} as described in Section~\ref{sec:methods}, as well as random selection. For \textsc{StratDegree} and \textsc{GreedyCover}, we split use the proposed algorithms to select 10\% of the data, stratified by class. The remaining 90\% of the data is randomly split (stratified by class) into validation (10\%) and training data (80\%). For random selection, we also want to determine whether adding more random training data improves classification robustness. Thus, in addition to using stratified random sampling to select 10\% of the data for training, we consider larger training sets, increasing to 30\% in 5\% increments. In all cases, 10\% of the data are used for validation and the remainder comprise the test set. We measure the average number of neighbors connected to a node outside of the training set, i.e., for the training set $T\subset V$, we record
\begin{equation}
    \frac{1}{\left|V\setminus T\right|}\sum_{i\in T}\sum_{j\in V\setminus T}{a_{ij}}.\label{eq:avgLabeledNeighbors}
\end{equation} 
This allows us to evaluate what impact the overall number of connections to the training data has on performance, and whether performance with the proposed training data selection methods match any trend observed with random training.

\subsection{Attacks}
\begin{itemize}
    \item\textbf{Nettack} The method from~\cite{Zugner2018}, briefly described in Section~\ref{sec:model}.
    \item\textbf{Fast Gradient Attack (FGA)} Computes the gradient of the loss function at the target node with respect to the adjacency matrix, then perturb the entry with the largest gradient that points in the correct direction~\cite{Chen2018}.
    \item\textbf{Integrated Gradient Attack (IG-Attack)}  A similar method that integrates the gradient as an entry in the adjacency matrix varies from 1 to 0 (for edge removal) or 0 to 1 (for edge addition)~\cite{WuH2019}.
    \item\textbf{Simplified Gradient Attack (SGA)} In this case, gradients are computed that only consider a $k$-hop subgraph around the target~\cite{Li2021}.
\end{itemize}
For direct attacks, we use up to 20 edge additions and removals for a target. For influence attacks, we allow up to 50 perturbations.

\subsection{Classifiers}
\label{subsec:classifiers}
We consider the following eight classifier models, some of which were developed with the explicit intent of improving robustness to adversarial attack:
\begin{itemize}
    \item\textbf{GCN} The original GCN architecture as used in~\cite{Zugner2018}.
    \item\textbf{Jaccard} Before training the GCN, removes edges between nodes that have dissimilar feature vectors before~\cite{Wu2019}.
    \item\textbf{SVD} Uses a GCN in which the adjacency matrix is replaced with a low-rank approximation via truncated singular value decomposition~\cite{Entezari2020}.
    \item\textbf{ChebNet} Uses the spectral graph convolutions~\cite{Defferrard2016} of which the convolution operator (\ref{eq:graphConv}) is a first-order approximation.
    \item\textbf{Simple Graph Convolution (SGC)} Applies a model similar to the surrogate (\ref{eq:surrogate}), where the matrix $W$ is learned via logistic regression on the features defined by $(D^{-1/2}AD^{-1/2})^kX$~\cite{WuF2019}.
    \item\textbf{Graph Attention Network (GAT)} Includes an attention mechanism based on the importance of each node's neighbors' features~\cite{Velivckovic2018}.
    \item\textbf{Robust Graph Convolutional Network (RGCN)} Uses Gaussian convolutions, in which the output is drawn from a Gaussian distribution whose parameters the output of a neural network~\cite{Zhu2019}.
    \item\textbf{MedianGCN} Aggregates neighbors' features based on their median values rather than weighted averages~\cite{Chen2021}.
\end{itemize}

\subsection{Training}
\label{subsec:training}
We tuned classifier hyperparameters for each (classifier, attack, training selection method) triple, first performing a coarse grid search over all hyperparameters, then performing some refinements: altering each single parameter 10\% and choosing the configuration with the best performance. The performance metric is a linear combination of the $F_1$ score (macro averaged) before an attack takes place with the was the average margin of 10 randomly selected targets after 5 perturbations with a direct attack. The resulting hyperparameters were used in all cases with the corresponding classifier, attack, and training selection method.

\subsection{Evaluation}
\label{subsec:eval}
We evaluate performance based on 25 target nodes. The targets are randomly selected from the set of nodes that are correctly classified when no attack takes place. This procedure is repeated five times with the train/validation/test splits recomputed each time. Our robustness metric is the adversary's required budget to achieve a given attack success rate. We compute this based on the number of perturbations required to give a target a negative classification margin in its correct class. If the target is never successfully misclassified, we set the required budget to the maximum number of perturbations. The result is averaged across the five trials.
\section{Results on Real Data}
\label{sec:results}
\subsection{Impact of Training Methods}
\label{subsec:varyTrain}

\begin{figure}
    \centering
    \includegraphics[width=0.8\textwidth]{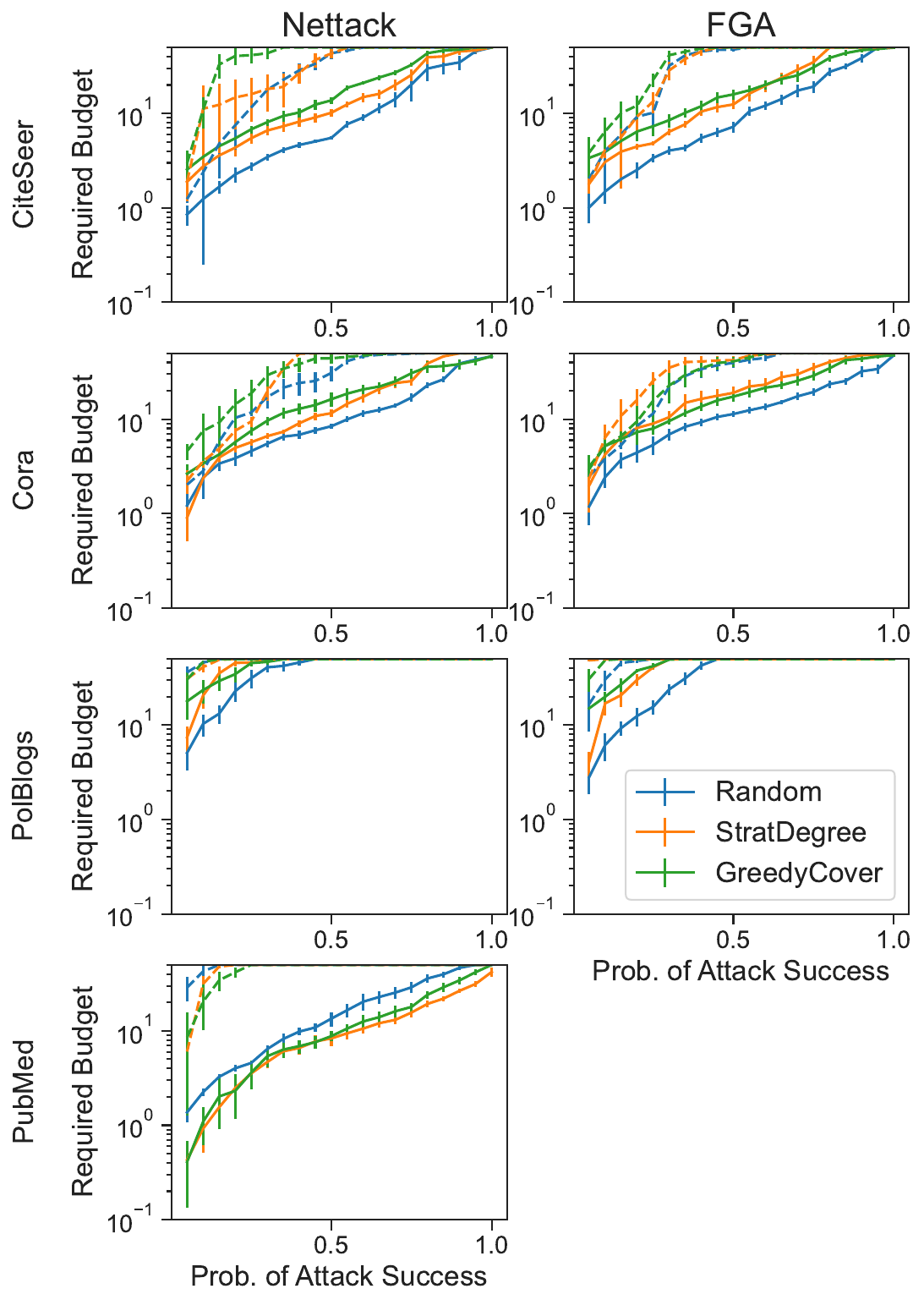}
    \caption{Robustness to influence attacks using GCNs (solid line) or with the best defense at a given attack success probability (dash line). Results are shown for the CiteSeer, Cora, PolBlogs, and PubMed datasets, each plotted in a subsequent row, and using both the Nettack/SGA (left column) and FGA (right column) attacks. Insufficient results were returned in the allotted time for IG-FGSM on all datasets, and FGA for PubMed. Each curve represents the average required budget over 25 randomly selected targets, and error bars are standard errors. Higher is better for the defender. With the exception of the PubMed dataset, \textsc{GreedyCover} performs at least as well as random training selection, and often performs much better.}
    \label{fig:indirect}
\end{figure}

We first consider influence attacks, where the target node's neighbors are modified rather than the target itself. We apply both Nettack and FGA, replacing Nettack with SGA if the SGC-based classifier is used. We only obtained results using IG-FGSM on the PolBlogs dataset, which can be seen in the appendix. (IG-FGSM did not substantially outperform the other methods for the best-performing classifiers.) In all other cases, IG-FGSM did not finish in the allotted time. Results are illustrated in Figure~\ref{fig:indirect}. In addition to standard the results for standard GCNs, we plot the upper envelope for each method: at a given attack success probability, the largest required budget across all classifiers. See Appendix~\ref{sec:realDataTables} for details about the performance of each individual classifier with each training scheme. CiteSeer has a particularly large increase in the attacker's required budget when using \textsc{GreedyCover}: more than doubling it over several rates of attack success. In fact, at low attack success probabilities, \textsc{GreedyCover} with a GCN provides similar robustness to any of the classifiers listed in Section~\ref{subsec:classifiers} with random selection. In addition, \textsc{GreedyCover} provides greater robustness when used in conjunction with the most robust defenses, as shown by the upper envelope. There is a somewhat milder effect on the Cora dataset. In this case, \textsc{GreedyCover} still performs best when using Nettack, but the best performance when attacked with FGA comes from \textsc{StratDegree} (though \textsc{GreedyCover} is within one standard error). With PolBlogs, we also see a benefit from both methods, though we start from a much higher baseline in terms of required perturbations. We see an exception with PubMed, where random training performs best. Looking deeper into the data, we see that the target nodes for random data tend to have higher margins on the best-performing classifiers. In all other cases, \textsc{GreedyCover} performs as well or better than the other training set selection methods.

\begin{observation}
Training with \textsc{GreedyCover} frequently outperforms other training methods, both with GCNs and in conjunction with published defenses.
\end{observation}

\begin{figure}
    \centering
    \includegraphics[width=0.9\textwidth]{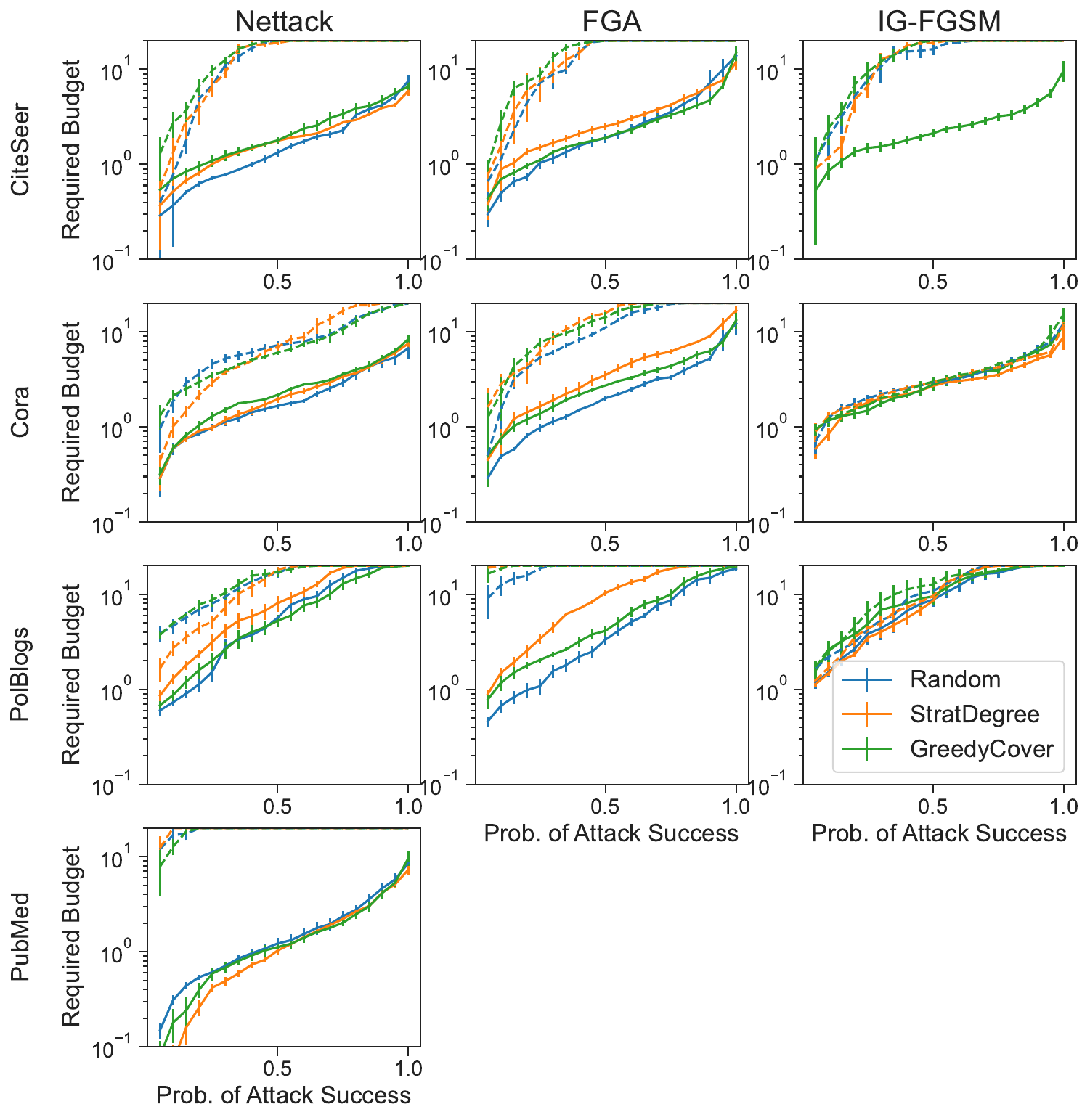}
    \caption{Robustness to direct attacks using GCNs (solid line) or with the best defense at a given attack success probability (dash line). Results are shown for the CiteSeer, Cora, PolBlogs, and PubMed datasets, attacked with Nettack/SGA (left column), FGA (center column), and IG-FGSM (right column). Insufficient results were returned in the allotted time for IG-FGSM and FGA on the PubMed dataset, or for IG-FGSM on the CiteSeer dataset when using a GCN with random training or \textsc{StratDegree}. Each curve represents the average required budget over 25 randomly selected targets, and error bars are standard errors. Higher is better for the defender. While \textsc{GreedyCover} performs better when paired with defenses on CiteSeer when attacked with Nettack or FGA, the alternative methods generally increase robustness less than with indirect attacks.}
    \label{fig:direct}
\end{figure}

When considering direct attacks, we use all four attacks, again with SGA replacing Nettack in the appropriate case. Results of these experiments are shown in Figure~\ref{fig:direct}. It is much more difficult to defend against direct attacks; note that the attacker often only needs one or two perturbations to be successful. With the CiteSeer dataset, we once again see higher robustness with \textsc{GreedyCover} and \textsc{StratDegree}, in particular at low attack probabilities. With Nettack and FGA, \textsc{GreedyCover} improves performance when combined with other defenses.  With IG-FGSM, on the other hand, the alternative training methods provide little benefit. Across datasets, this attack also has the lowest robustness across defenses, which would suggest it is a preferred attack for adversaries. (Note that for CiteSeer, we only obtained data for a GCN classifier when attacked with IG-FGSM when using \textsc{GreedyCover}.) We once again see a detriment in performance with PubMed, though in this case in the area where a perturbing a single edge results in an successful attack.

\begin{observation}
    Direct attacks typically benefit less than influence attacks from the alternative training methods.
\end{observation}


\subsection{Impact of Labeled Neighbors}
\begin{figure}
    \centering
    \includegraphics[width=0.95\textwidth]{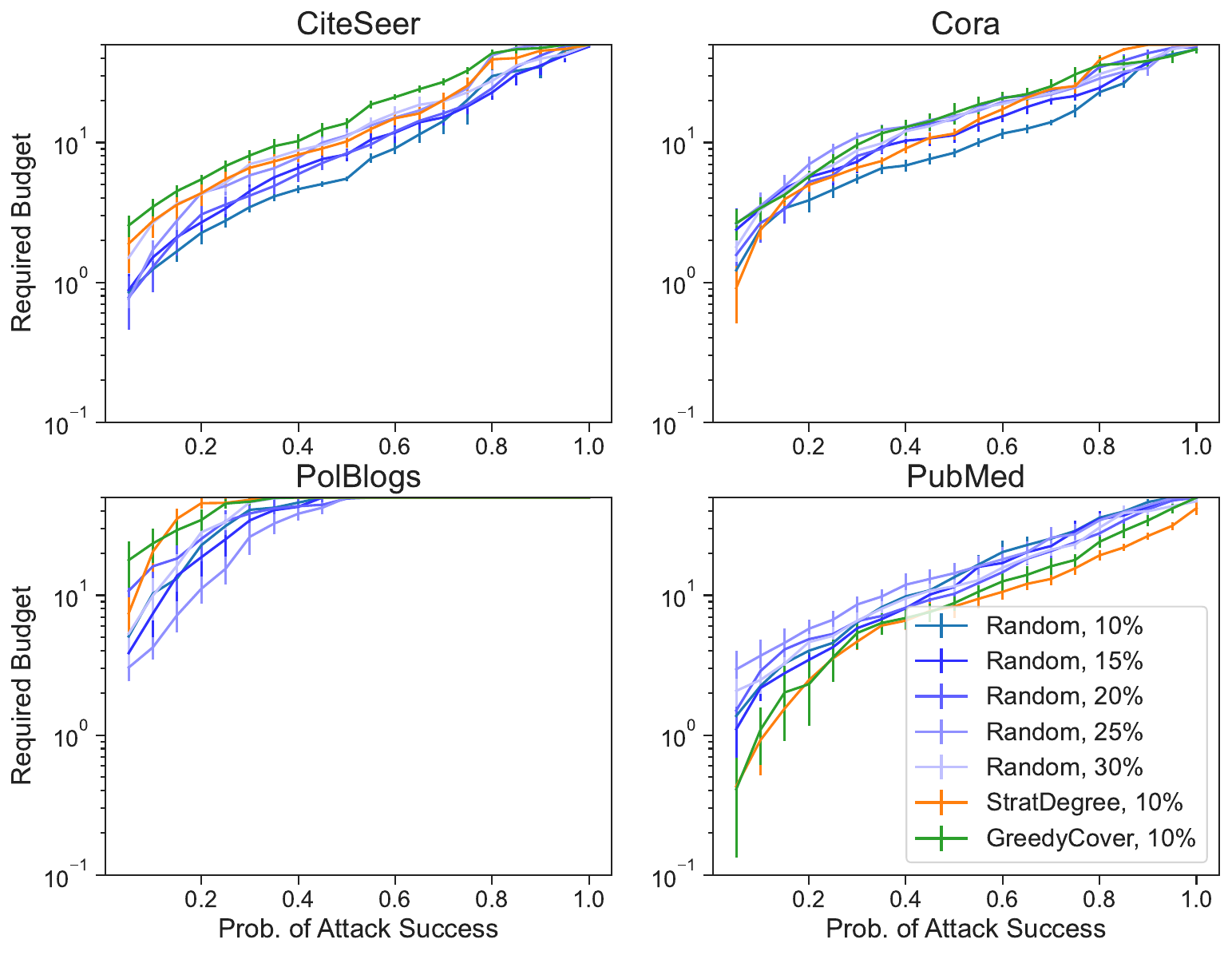}
    \caption{Robustness to influence attacks using GCNs when training data are selected using \textsc{GreedyCover}, \textsc{StratDegree}, or varying amounts of random selection. Results are shown for the CiteSeer (upper left), Cora (upper right), PolBlogs (lower left), and PubMed (lower right) datasets. Each curve represents the average required budget over 25 randomly selected targets, and error bars are standard errors. Higher is better for the defender. Of the datasets where robustness improves using \textsc{GreedyCover} (i.e., CiteSeer, Cora, and PolBlogs), the only case that consistently performs better than \textsc{GreedyCover} is 30\% random selection on the Cora dataset.}
    \label{fig:moreTraining}
\end{figure}

One possibility we considered is that robustness from the alternative training methods comes entirely from the average number of trained neighbors for nodes in the test set. To test this possibility, we performed the same experiments with more randomly selected training data, as described in Section~\ref{subsec:dataselection}. Results of these experiments are shown in Figure~\ref{fig:moreTraining}, using Nettack as an influence attack against a GCN. While using more randomly selected training data does sometimes increase robustness, it is not consistent, and in some cases more randomly selected training data is slightly less robust. The one case where additional training data consistently outperforms \textsc{GreedyCover} in terms of robustness is Cora, where training using 30\% of the dataset, randomly selected, outperforms the alternatives. In this case, the average number of neighbors per target for \textsc{GreedyCover} and \textsc{StratDegree} are 1.084 and 1.135, both between the values for 25\% random training (1.029) and 30\% (1.237). Thus, increasing the number of neighbors in the training set by adding more randomly selected training data does not necessarily increase classifier robustness to the same extent.

\begin{observation}
    Using more training data with random selection does not consistently lead to higher robustness.
\end{observation}

\subsection{Robustness vs. Classification Performance}
\begin{figure}
    \centering
    \includegraphics[width=0.95\textwidth]{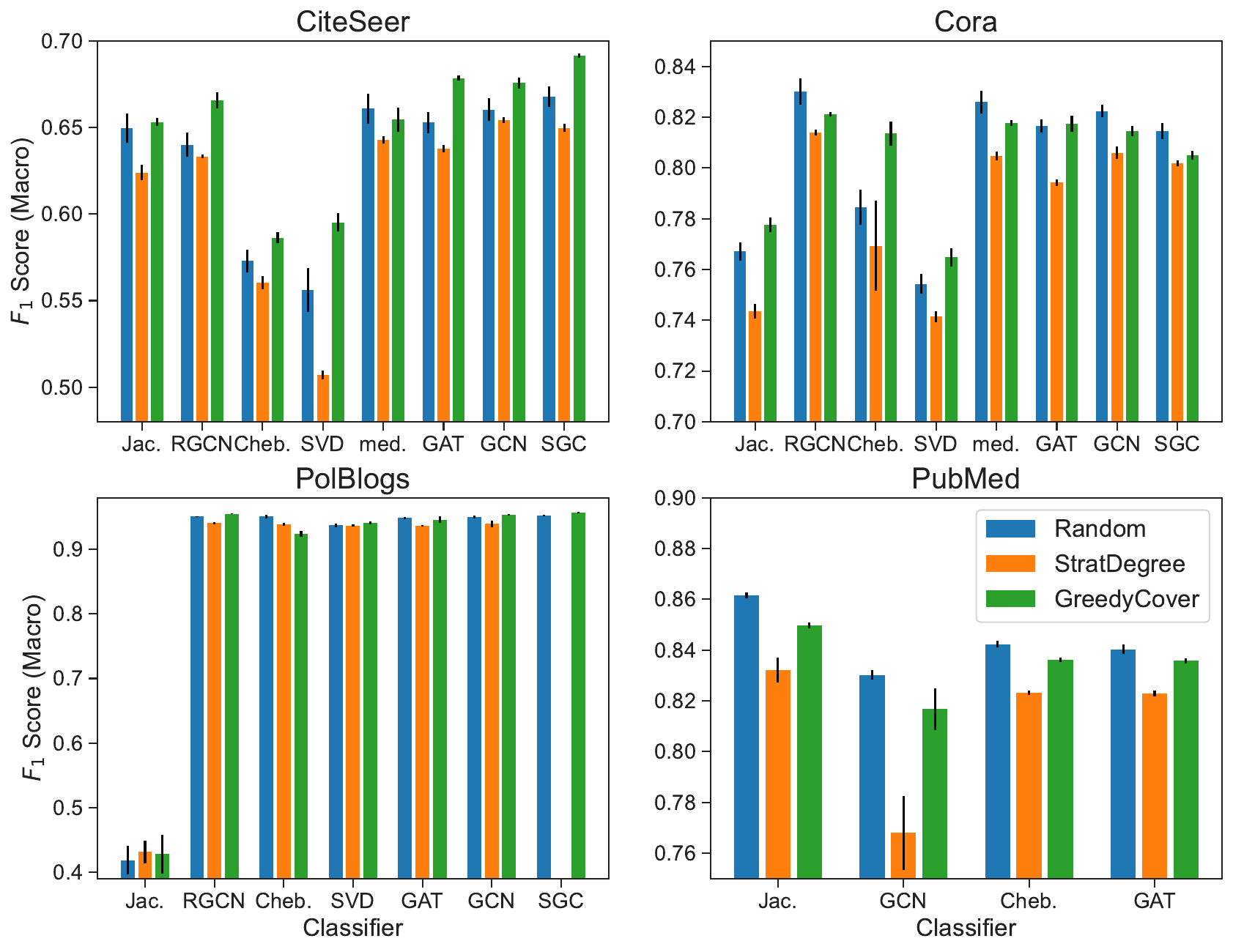}
    \caption{Classifier performance across datasets when training data are selected using \textsc{GreedyCover}, \textsc{StratDegree}, or varying amounts of random selection. Results are shown for the CiteSeer (upper left), Cora (upper right), PolBlogs (lower left), and PubMed (lower right) datasets. Each bar height represents the average $F_1$ score (macro averaged) across 5 separate train/validation/test sets, and error bars are standard errors. Performance is shown for each classifier where experiments completed within the allotted time. Higher is better for the defender. While \textsc{StratDegree} often underperforms random selection, \textsc{GreedyCover} typically shows similar performance.}
    \label{fig:classification}
\end{figure}

Another important consideration is whether increased robustness comes at the expense of classification performance. In Figure~\ref{fig:classification}, we show the macro-averaged $F_1$ score for each method using all classifiers. Performance does occasionally vary. In particular, \textsc{StratDegree} results in somewhat lower performance than random training among most classifiers for all datasets. \textsc{GreedyCover}, on the other hand, typically yields similar performance to random selection and occasionally outperforms it, e.g., using SGC on CiteSeer and ChebNet on Cora. This yields another datapoint in favor of \textsc{GreedyCover}: it tends to yield the greatest robustness across datasets, and does not seem to greatly hinder overall classification performance.

\begin{observation}
    Using \textsc{GreedyCover} yields no consistent reduction in classification performance compared to random training set selection.
\end{observation}

\subsection{Adaptive Attacks}
In the image classification literature, numerous published defenses were found to primarily rely on model obfuscation and remain vulnerable to adaptive attacks that take the new model into account~\cite{Athalye2018}. Recent work has raised similar concerns regarding the robustness of published defenses against GNN attacks~\cite{Mujkanovic2022}. If training set selection makes a classifier more robust, one advantage is that it makes no changes to the model class, and thus should not be vulnerable to such oversights.

\begin{figure}
    \centering
    \includegraphics[width=0.5\textwidth]{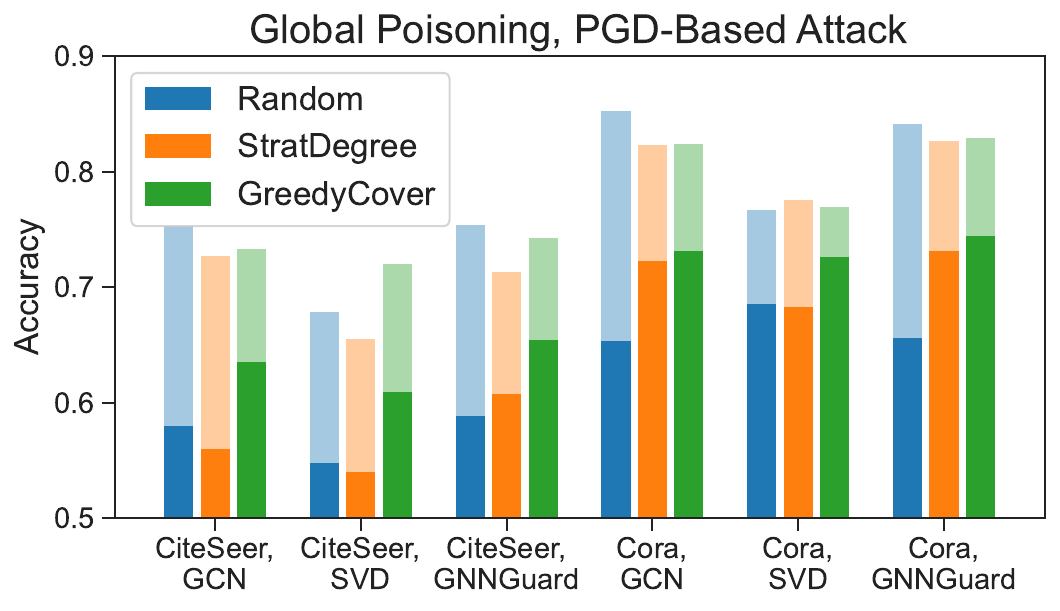}
    \caption{Performance using an adaptive attack for global poisoning with all three training schemes. Results are shown in terms of overall classifier accuracy using a GCN, an SVD-based GCN, and GNNGuard on the CiteSeer and Cora datasets. Bars showing accuracy before poisoning is desaturated, while accuracy after poisoning is solid. Higher accuracy is better for the defender. In all cases, selecting training data using \textsc{GreedyCover} results in better post-attack accuracy.}
    \label{fig:adaptive_result}
\end{figure}

We applied our training set selection methods to a demonstration provided by Muj\-kanovic and Geisler et al.\footnote{Available at \url{https://github.com/LoadingByte/are-gnn-defenses-robust}.}, which includes an adaptive attack based on projected gradient descent (PGD)~\cite{Xu2019}. The code applies the attack with the objective of reducing the overall classifier accuracy. We applied the demo to the same datasets used by the authors---CiteSeer and Cora---and achieved the results shown in Figure~\ref{fig:adaptive_result}. While the SVD-based method and GNNGuard are both effectively attacked by the PGD-based method, using \textsc{GreedyCover} to select the training data (again using 10\% for training, 10\% for validation, and 80\% for testing) results in higher post-attack accuracy for with both classifiers. As defenses to new adaptive methods are published, it will be interesting to consider their use in conjunction with alternative training set selection.

\section{Simulation Study}
\label{sec:simulation}

The results on real data show that \textsc{GreedyCover} often provides greater robustness to attack, but they are by no means conclusive. In the next section, we further explore the methods with simulated data to observe performance differences while controlling network properties.

\subsection{Synthetic Dataset Generation}
Synthetic network generation to evaluate network effects on the performance of GNNs has recently received attention in the research community~\cite{Palowitch2022}. In this section, we outline a set of simulations that consider four key features of the network: degree distribution, level of clustering, homophily with respect to labels, and information gained via node attributes.
\subsubsection{Random Graph Models}
We use five random graph models that exhibit different properties in terms of clustering, degree distribution, and dependence on attributes. In each case, we use 1200 nodes and an average degree of approximately 10.
\begin{itemize}
    \item \textbf{Erd\H{o}s--R\'{e}nyi (ER) Graphs}: Each pair of nodes shares an edge with probability $1/120$. This model yields homogeneous degree distributions and very little clustering.
    \item \textbf{Barab\'{a}si--Albert (BA) Graphs}: Each node enters the graph and connects 5 edges to existing nodes with probability proportional to their degrees. The process is initialized with a 6-node star. This model yields graphs with heterogeneous degree distributions and very little clustering.
    \item \textbf{Watts--Strogatz (WS) Graphs}: A ring lattice---where each node is connected to 5 nodes on either side---has 10\% of its edges randomly rewired. This model yields graphs with substantial clustering and homogeneous degree distributions.
    \item \textbf{Lancichinetti--Fortunato--Radicchi (LFR) Graphs}: Generates a degree sequence with degree distribution $p(d)\propto d^{-3},$ with average and minimum degree set to $d_\textrm{avg}=10$ and $d_\textrm{max}=135$. Nodes are randomly assigned to communities, whose sizes are distributed according to $p(|C|)\propto|C|^{-2}$, with the minimum community size being 10. Nodes create 80\% of their connections within the community and 20\% outside the community.
\end{itemize}
If the generated graph has multiple connected components, we use the largest connected component for the experiment.

\subsubsection{Label Assignment}
\label{subsubsec:labels}
We assign labels with varying levels of homophily. For the ``high homophily'' scenario, we partition the nodes based on the normalized graph Laplacian $$L=I-D^{-1/2}AD^{-1/2},$$ where $A$ and $D$ are the adjacency matrix and diagonal degree matrix as in Section~\ref{sec:model}~\cite{Chung1997}. We select the eigenvector $u_2$ associated with the second-smallest eigenvalue of $L$. The nodes associated with the $N/2$ entries in $u_2$ with the smallest values (i.e., values closest to $-\infty$) are labeled 0, and the other nodes are labeled 1. Let $V_0$ and $V_1$ be the respective subsets of vertices. 

For lower homophily graphs, we first compute the difference between the number of within-label edges and the number of cross-label edges, i.e., letting $E_{ij}$ be the set of edges between nodes in $V_i$ and nodes in $V_j$,
\begin{equation}
\Delta=|E_{00}|+|E_{11}|-|E_{01}|.
\end{equation}
Depending on how homophilous we want the graph to be, we swap labels on pairs of nodes until $\Delta$ reaches a given value, based on its value from the initial Laplacian-based partition (e.g., half as homophilous as the original). The node swapping mechanism is detailed in Algorithm~\ref{alg:swap}.

\begin{algorithm}
\begin{centering}
\begin{algorithmic}
\STATE \textbf{Input:} Graph $G=(V,E)$, vertex subsets $V_0, V_1$
\STATE \textbf{Output:} Updated vertex subsets $\hat{V}_0, \hat{V}_1$
\STATE $x_i\gets\begin{cases}-1 & \textrm{if }v_i\in V_0\\
1 & \textrm{if }v_i\in V_1\end{cases}$
\STATE $d \gets Ax$
\STATE $p_i\gets d_i\cdot\mathbb{I}[d_i >= 0]\cdot\mathbb{I}[x_i > 0]$
\STATE $u\gets$ node randomly selected with probability $p_i/\sum_{j=1}^N{p_j}$
\STATE $p_i\gets d_i \mathbb{I}[d_i < 0]\mathbb{I}[x_i < 0]$
\STATE $v\gets$ node randomly selected with probability $p_i/\sum_{j=1}^N{p_j}$
\STATE $\hat{V}_0\gets (V_0\cup\{u\})\setminus{v}$
\STATE $\hat{V}_1\gets (V_1\cup\{v\})\setminus{u}$
\RETURN $\hat{V}_0, \hat{V}_1$
\end{algorithmic}
\caption{\textsc{Swap2Reduce}}
\label{alg:swap}
\end{centering}
\end{algorithm}
\subsubsection{Synthetic Attributes}
\label{subsubsec:attributes}
As with the real graphs, we consider binary attribute vectors on the nodes of the synthetic graphs. In each case, we consider nodes with 100 attributes, and we give each attribute a probability of being true depending on its label. Probabilities are determined by an exponentially decreasing function. We consider three scenarios. In the most difficult case, the probabilities are the same for both classes. As we make the problem easier, we shift the  function that determines the attribute probabilities so that high-probability attributes in class 0 still have \emph{relatively} high probabilities class 0, but there is not a perfect match. 
The shifts were chosen to create cases where a generalized likelihood ratio test (with each attribute having an independent probability parameter, estimated based on 60 cases for each class) achieves accuracy of approximately $0.5$, $0.7$, and $0.9$. We refer to these cases as having \emph{uninformative}, \emph{moderately informative}, and \emph{highly informative} attributes, respectively. In addition, each node has a one-hot encoded attribute indicating its index in the node set.

\subsection{Results}
For all synthetic topologies, we ran experiments using Nettack perform an influence attack against a GCN trained with data selected by all three methods. Robustness results for ER, BA, WS, and LFR graphs are shown in Figure~\ref{fig:sim_budget}. When no attributes are used and homophily is high, we see a much larger performance difference using \textsc{GreedyCover} than \textsc{StratDegree} in the WS graphs, but the two methods yield more similar performance with the other models. For all models, as homophily decreases, the performance improvement gets more modest as homophily decreases.

 When attributes with the same distribution are added to both classes (i.e., the case of ``uninformative'' attributes), robustness suffers in most cases. The LFR graphs in particular see a large decrease in robustness using random selection, with a much smaller decrease using the alternative methods. As feature distributions become more distinct between the classes, the difference between the methods becomes smaller, suggesting that the robustness improvements we observe are likely due to structural considerations. With highly informative attributes, we note that the models with homogeneous degree distributions still gain a benefit from \textsc{StratDegree} and \textsc{GreedyCover} when homophily is low, while the models with heterogeneous degree distributions are somewhat hindered by these methods. Like in the real data, this is because the targets have higher margins in the case of random training selection. This may happen due to low-degree nodes, which tend to connect to high-degree nodes: When homophily is low, nodes may become more difficult to predict based on their proximity to hubs, and are \emph{less} likely to be selected to be labeled. We summarize our observations here as follows:
\begin{figure}
    \centering
    \includegraphics[width=0.95\textwidth]{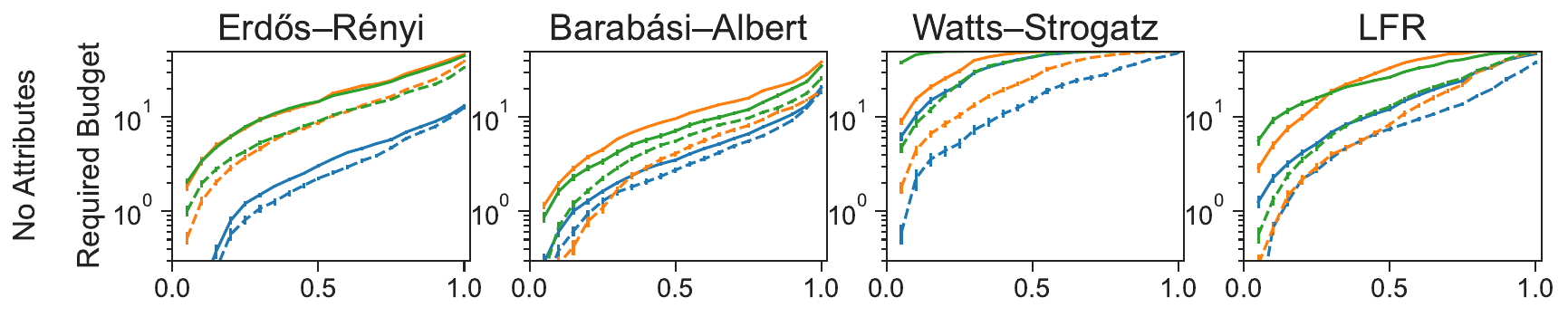}\\
    \includegraphics[width=0.95\textwidth]{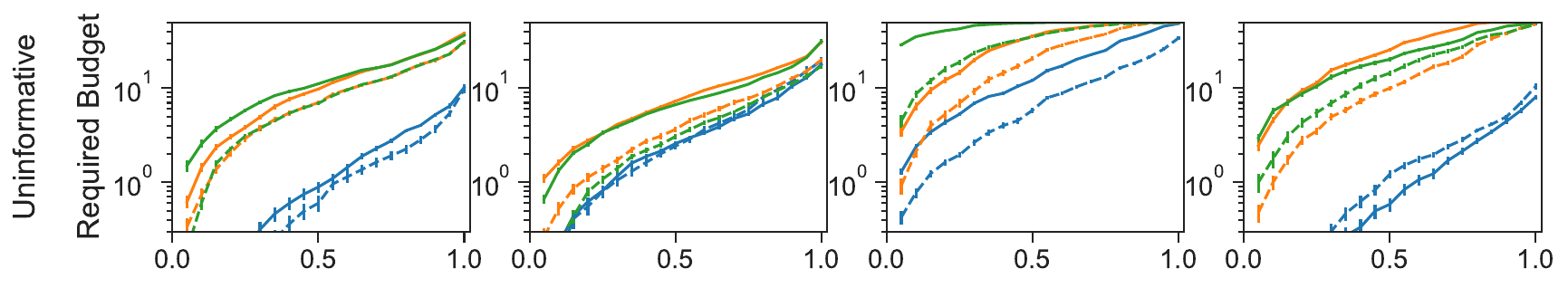}\\
    \includegraphics[width=0.95\textwidth]{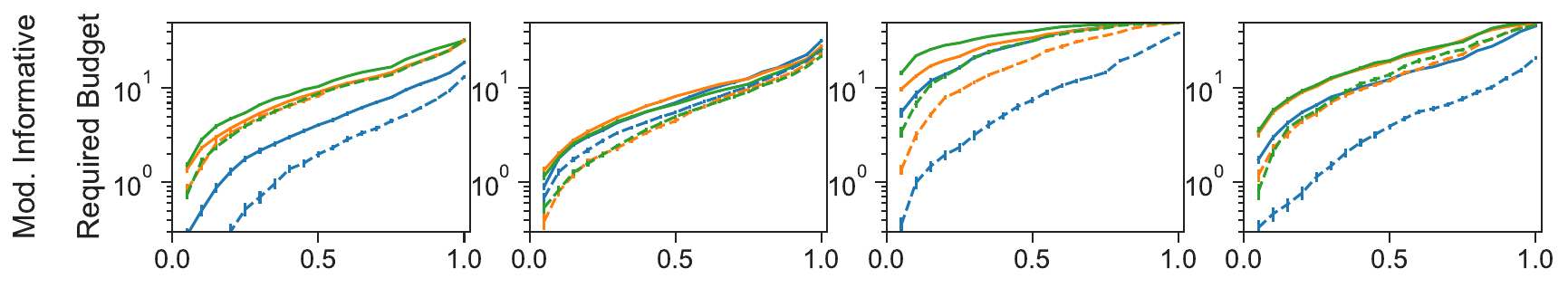}\\
    \includegraphics[width=0.95\textwidth]{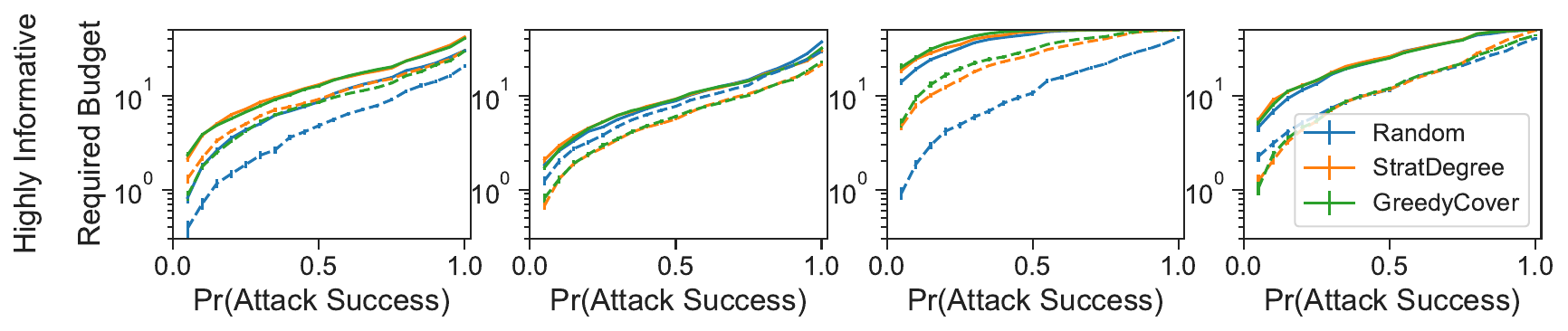}
    \caption{Robustness to influence attacks against GCNs on simulated data. Results are shown for ER (first column), BA (second column), WS (third column), and LFR (fourth column) graphs, in cases with no attributes (first row), uninformative attributes (second row), moderately informative attributes (third row), and highly informative attributes (fourth row).  Each curve represents the average required budget over 25 randomly selected targets, and error bars are standard errors. Higher is better for the defender. Results are shown for high homophily (solid line) and low homophily (dash line) cases. As attributes become more helpful in classification, the advantage gained by the alternative training methods is substantially reduced.}
    \label{fig:sim_budget}
\end{figure}
\begin{observation}
With no attributes and high homophily, all models gain robustness from the alternative methods.
\end{observation}
\begin{observation}
With no attributes and low homophily, \textsc{GreedyCover} provides robustness for all models, while for BA and LFR, \textsc{StratDegree} improves robustness only at higher attack success rates.
\end{observation}
\begin{observation}
The increase in robustness for the alternative methods decreases as homophily decreases and as attributes become better class predictors.
\end{observation}
\begin{observation}
With highly informative attributes and low homophily, \textsc{Gree\-dyCover} and \textsc{StratDegree} maintain some increased robustness for homogeneous degree distributions, while they somewhat hinder performance for heterogeneous ones.
\end{observation}

\begin{figure}
    \centering
    \includegraphics[width=0.95\textwidth]{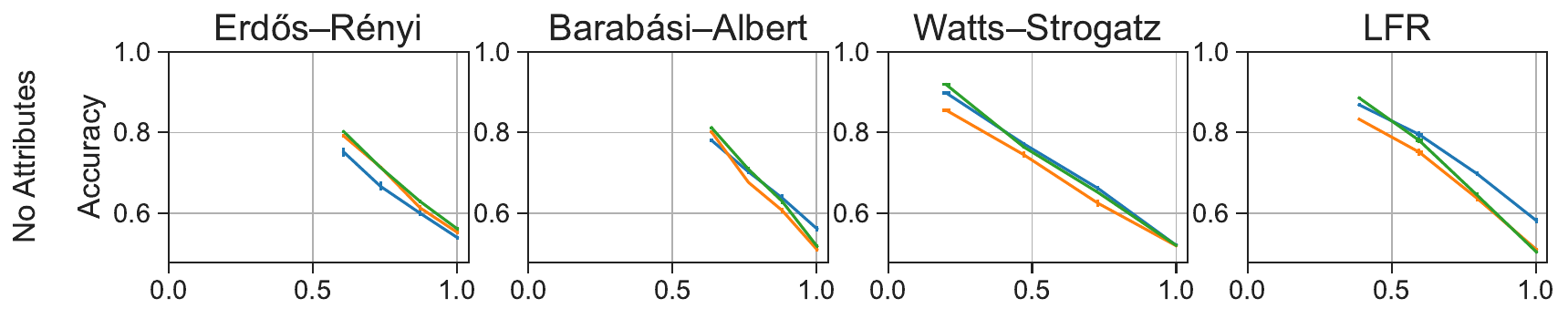}\\
    \includegraphics[width=0.95\textwidth]{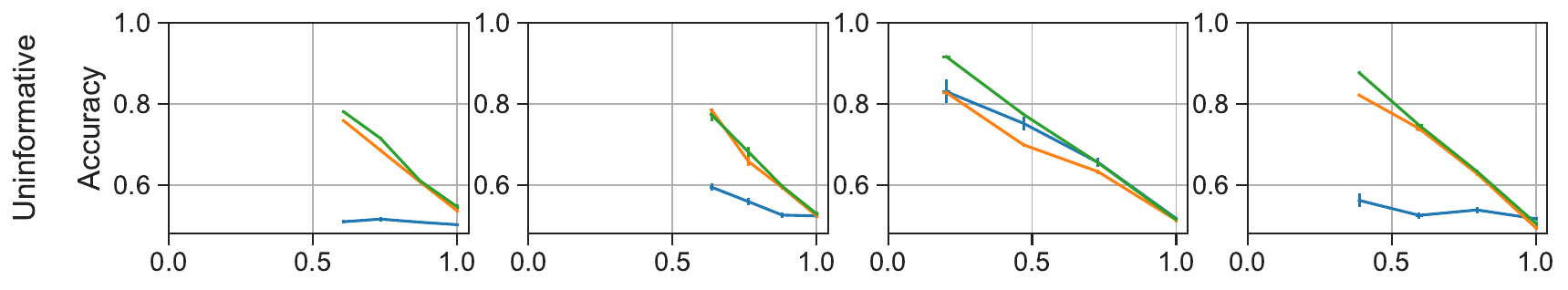}\\
    \includegraphics[width=0.95\textwidth]{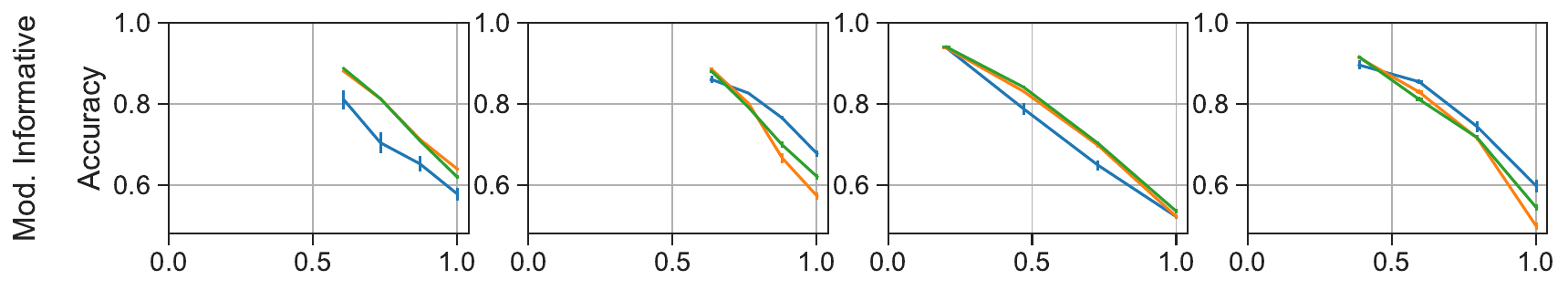}\\
    \includegraphics[width=0.95\textwidth]{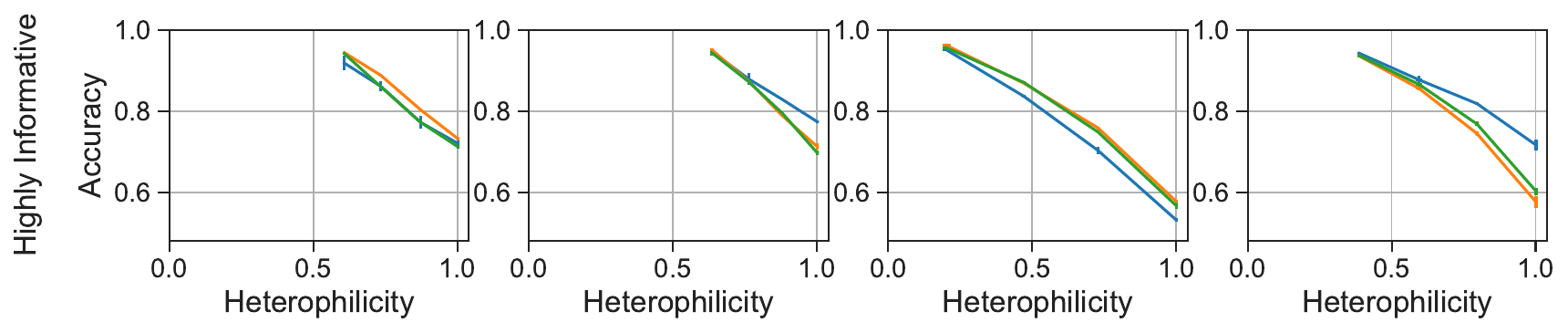}
    \caption{Classification accuracy as a function of heterophilicity using GCNs on simulated data. Results are shown for ER (first column), BA (second column), WS (third column), and LFR (fourth column) graphs, in cases with no attributes (first row), uninformative attributes (second row), moderately informative attributes (third row), and highly informative attributes (fourth row).  Each curve represents the average required budget over 5 train/validation/test splits, and error bars are standard errors. Higher is better for the defender. The principal performance differences occur with skewed degree distributions when homophily is low.}
    \label{fig:sim_accuracy}
\end{figure}

We consider the potential impact of the alternative training data on classifier performance as well. Results are shown in Figure~\ref{fig:sim_accuracy}. Since we use two balanced classes in all cases, we use accuracy as the classification metric. For each case, we plot accuracy as a function of  \emph{heterophilicity}~\cite{Park2007}, computed as 
\begin{equation}
H=\frac{\#\{\textrm{edges between $V_0$ and $V_1$}\}}{|V_0||V_1|M/\binom{N}{2}}.\label{eq:heterophilicity}
\end{equation}
The denominator in (\ref{eq:heterophilicity}) is the expected number of edges between $V_0$ and $V_1$ after random rewiring. A high-homophily graph will have relatively low heterophilicity. Note that both ER and BA graphs span the same range of heterophilicity, while LFR graph can achieve lower heterophilicity and WS can be almost perfectly homophilous. When no attributes are used, performance is similar across methods in the high-homophily (low-heterophilicity) cases, while the alternative methods perform worse in the low-homophily cases. This yields a significant gap in the in the cases with skewed degree distributions. In particular, LFR graphs maintain 75\% accuracy with random training even in the case where there is no homophily (heterophilicity is 1). As in the analogous results in Figure~\ref{fig:sim_budget}, this may be due to low-degree nodes that are unlikely to be chosen as training data, but are more difficult to classify in a less homogeneous setting.

As attributes are added to the graphs, we see a decrease in performance when the uninformative attributes are added, though the difference is very small using  \textsc{GreedyCover} for the clustered models. As we expect, accuracy increases as the attributes become more informative. As we observed in the robustness results, we see differences between methods diminish as attributes help discriminate the classes.

\begin{observation}
Graphs with skewed degree distributions and low homophily achieve lower accuracy with \textsc{GreedyCover} and \textsc{StratDegree} than random selection, but performance is similar in other cases.
\end{observation}
\begin{observation}
For higher homophily graphs, performance differences between methods decrease as attributes become more informative.
\end{observation}

When heterophilicity is approximately 1, accuracy is very low without informative attributes. Considering cases where there is at least some homophily and at least moderately informative attributes, the simulation results where robustness does not improve with \textsc{StratDegree} or \textsc{GreedyCover} are summarized in Table~\ref{tab:summary}. As shown in the table, the cases where there is no improvement all have heterogeneous degree distributions, while the homogeneous degree distributions always have some improvement in robustness when attributes are at least moderately informative. In addition, the low homophily cases result in lower accuracy with the alternative methods. Note also that more informative attributes and lower clustering coefficient hinder the performance benefit.
\begin{table}
\begin{centering}
\caption{Summary of cases where \textsc{StratDegree} and \textsc{GreedyCover} do not improve robustness when using informative attributes on synthetic graphs. The alternative methods are considered less robust than random training selection if the adversary's budget decreases by at least 1 standard deviation for at least 10 out of 20 points on the associated curve in Figure~\ref{fig:sim_budget} (attack success probability in multiples of 0.05). They are considered to be similarly robust if the budget is within 1 standard deviation of for over 10 such points. For accuracy, the alternative methods result in lower accuracy if the average accuracy in Figure~\ref{fig:sim_accuracy} decreases by at least 3\% and similar accuracy if it is within 3\%. All cases in the table have heterogeneous degree distributions. All cases with lower accuracy have low homophily. The improvement from the alternatives is also degraded as attributes become more informative (from 70\% to 90\% accuracy based on attributes alone) and clustering coefficient decreases.}
\label{tab:summary}
\includegraphics[width=\textwidth]{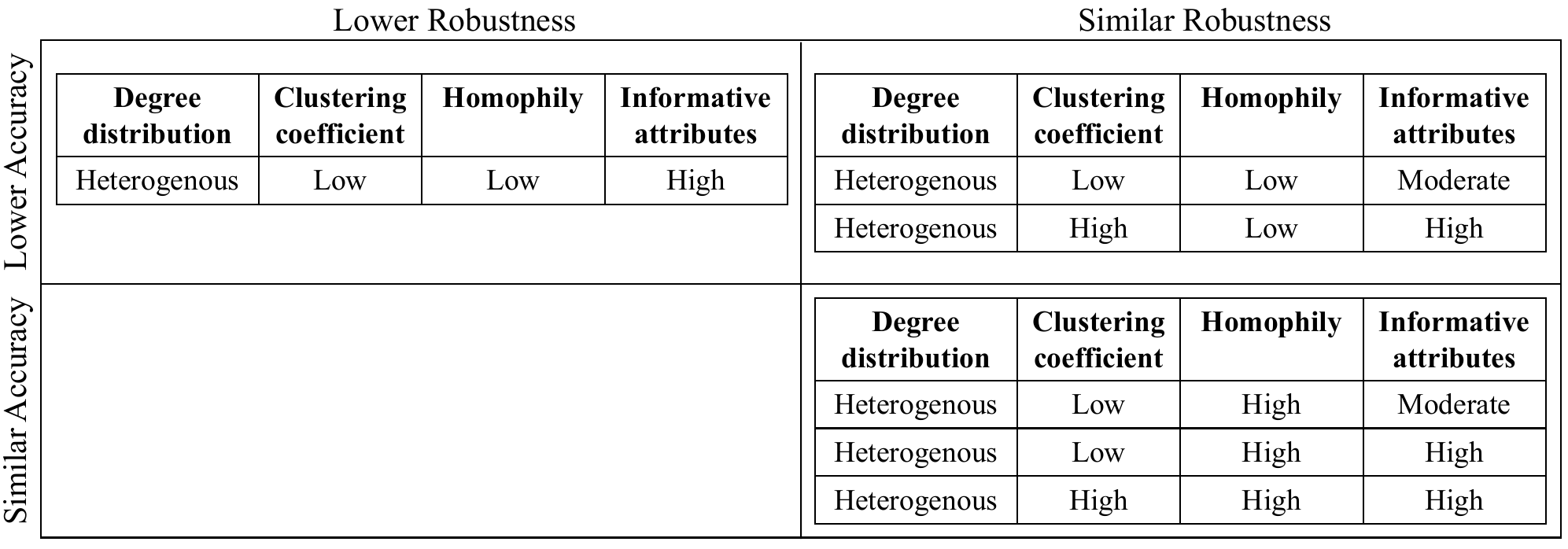}
\end{centering}
\end{table}

Relating the synthetic data to the real datasets, recall that \textsc{StratDegree} and \textsc{GreedyCover} both failed to provide consistent improvement for PubMed, and struggled with direct attacks against PolBlogs. Looking into the features of these datasets, two points of interest are that PolBlogs has an especially heavy-tailed degree distribution: there are many nodes with hundreds of edges, which is rare in the other datasets. In addition, the PubMed dataset has node attributes that are very useful in identifying the class of the nodes: a support vector machine with a radial basis function kernel trained on the attribute vectors alone (50\% of the nodes used for training), the $F_1$ score (macro averaged) for the Cora dataset is approximately 0.71, for CiteSeer is approximately 0.75, and for PubMed is about 0.87. As with the synthetic data, the cases with the most informative node attributes are hindered by the alternative training methods.

\section{Conclusions}
\label{sec:conclusion}
This paper explores the impact of complex network characteristics on the robustness of vertex classification using GCNs. In particular, we consider various scenarios regarding the structural relationship between the training data and the remainder of the network. We consider selecting training data using alternatives to random selection: using the highest degree nodes (\textsc{StratDegree}) and using nodes that result in more connections to training nodes from the test set (\textsc{GreedyCover}). We see the greatest improvement using \textsc{GreedyCover} against influence attacks, though there are improvements in other cases as well. With direct attacks via IG-Attack, on the other hand, performance is similar across methods, and robustness in the best case (or worst case for the defender) is lower than with the other attacks. We show that the robustness achieved against Nettack with the alternative training methods is not replicated through increasing the amount of randomly selected training data, and that there is no significant tradeoff between classifier performance and robustness using \textsc{GreedyCover}. In addition, we test \textsc{StratDegree} and \textsc{GreedyCover} against an adaptive global poisoning attack and show that \textsc{GreedyCover} yields better post-attack accuracy than random training.

In simulation, we see other interesting phenomena in the context of influence attacks: \textsc{GreedyCover} increases robustness against Nettack for a diverse set of topologies when label homophily is high and there are no node attributes. We find that \textsc{GreedyCover} and \textsc{StratDegree} cease to be helpful when homophily is very low and degree distributions are heterogeneous, perhaps because there are fewer labels on low-degree nodes that attach to hubs. In all cases, variation between training selection methods becomes less pronounced as node attributes become more helpful in discriminating between classes.

The work documented here points to several open problems and avenues of potential investigation. 
First, it is interesting that the integrated gradient method is consistently the strongest attack against real data, regardless of how training data are selected. Determining whether some network phenomenon can be exploited to improve robustness against these attacks would be an interesting topic for future work. Considering additional models for topologies and attributes could yield additional insight into where the various methods perform best, with Google's GraphWorld being an important enabling technology~\cite{Palowitch2022}. Another interesting question is whether there are certain topology--attribute combinations where there is a true tradeoff between robustness and classification performance. Identifying such cases---in the spirit of~\cite{Tsipras2019}, focused on graph data---would be important to understand what could make classification inherently vulnerable to attack. Another potential area to consider is detectability. Attackers try to hide their manipulation of the data; what would be necessary to determine that an attack has been performed on a graph? For example, we observed that direct attacks from Nettack increase triangle count~\cite{Miller2019}. There may be other network statistics that tend to change when an attack is carried out. These are all interesting questions to consider as the research community continues to expand its knowledge of vulnerability and robustness in graph machine learning.

\section*{Acknowledgement}
The authors wish to thank Mustafa \c{C}amurcu and Alexander J. Gomez for their assistance with the initial experiments on \textsc{StratDegree} and \textsc{GreedyCover}.
\bibliographystyle{plain}
\bibliography{bibfile}
\appendix
\section{Extended Experimental Results}
\label{sec:realDataTables}
We present results for each classifier at various attacker budgets, and highlight the best-performing pairing of a classifier with a training data selection method. Results on influence attacks for CiteSeer, Cora, Polblogs, and PubMed are in Tables~\ref{tab:citeseer_indirect}, \ref{tab:cora_indirect}, \ref{tab:polblogs_indirect}, and~\ref{tab:pubmed_indirect}, respectively. Results for direct attacks are likewise in Tables~\ref{tab:citeseer_direct}, Tables~\ref{tab:cora_direct}, Tables~\ref{tab:polblogs_direct}, Tables~\ref{tab:pubmed_direct}. As we saw in Section~\ref{subsec:varyTrain}, \textsc{GreedyCover} and \textsc{StratDegree} perform best for CiteSeer and Cora, and for PolBlogs in the case of influence attacks.
\begin{table}[h!]
\caption{Results of influence attacks against each classifier with the CiteSeer dataset, with attacker budgets of 10, 30, and 50 edge perturbations. Results are included for Nettack (Net), FGA, and IG-FGSM (IG). For each classifier, we train with random (Rand.), \textsc{StratDegree} (SD), and \textsc{GreedyCover} (GC). Each entry is a probability of attack success, thus higher is better for the attacker and lower is better for the defender. To yield the most robust classifier, the defender picks the classifier/training method combination that minimizes the worst-case attack probability. These entries are listed in \textbf{bold}. Entries representing the most robust case for random training are in \textit{italic}. Entries listed as N/A did not finish in the allotted time. The Jaccard-based classifier performs best, both overall (with \textsc{GreedyCover}) and using random training.}
\label{tab:citeseer_indirect}
\footnotesize
  \begin{tabular}{lc|ccc|ccc|ccc}
\hline
&&\multicolumn{3}{c|}{Budget 10}
&\multicolumn{3}{c|}{Budget 30}
&\multicolumn{3}{c}{Budget 50}\\
defense&train.&Net&FGA&IG&Net&FGA&IG&Net&FGA&IG\\
\hline
Jaccard&Rand.
&\textit{0.248}
&0.224
&N/A
&\textit{0.432}
&0.336
&N/A
&\textit{0.504}
&0.408
&N/A
\\
Jaccard&SD
&0.28
&0.24
&N/A
&0.4
&0.36
&N/A
&0.448
&0.368
&N/A
\\
Jaccard&GC
&0.152
&\textbf{0.168}
&N/A
&0.176
&\textbf{0.288}
&N/A
&0.192
&\textbf{0.36}
&N/A
\\
RGCN&Rand.
&0.752
&0.696
&N/A
&0.944
&0.84
&N/A
&0.976
&0.904
&N/A
\\
RGCN&SD
&0.504
&0.328
&N/A
&0.768
&0.592
&N/A
&0.816
&0.744
&N/A
\\
RGCN&GC
&0.448
&0.384
&N/A
&0.864
&0.76
&N/A
&0.952
&0.824
&N/A
\\
Cheb&Rand.
&0.304
&0.384
&N/A
&0.496
&0.464
&N/A
&0.552
&0.544
&N/A
\\
Cheb&SD
&0.32
&0.352
&N/A
&0.464
&0.432
&N/A
&0.52
&0.496
&N/A
\\
Cheb&GC
&0.352
&0.272
&N/A
&0.52
&0.424
&N/A
&0.552
&0.512
&N/A
\\
SVD&Rand.
&0.536
&0.424
&N/A
&0.816
&0.696
&N/A
&0.904
&0.88
&N/A
\\
SVD&SD
&0.624
&0.6
&N/A
&0.912
&0.776
&N/A
&0.968
&0.928
&N/A
\\
SVD&GC
&0.344
&0.376
&N/A
&0.624
&0.6
&N/A
&0.808
&0.688
&N/A
\\
median&Rand.
&0.584
&0.44
&N/A
&0.816
&0.84
&N/A
&0.864
&0.88
&N/A
\\
median&SD
&0.424
&0.376
&N/A
&0.8
&0.784
&N/A
&0.88
&0.912
&N/A
\\
median&GC
&0.392
&0.304
&N/A
&0.768
&0.768
&N/A
&0.88
&0.864
&N/A
\\
GAT&Rand.
&0.624
&0.568
&N/A
&0.928
&0.848
&N/A
&0.976
&0.936
&N/A
\\
GAT&SD
&0.552
&0.392
&N/A
&0.816
&0.728
&N/A
&0.912
&0.896
&N/A
\\
GAT&GC
&0.424
&0.328
&N/A
&0.864
&0.752
&N/A
&0.936
&0.888
&N/A
\\
GCN&Rand.
&0.68
&0.568
&N/A
&0.848
&0.856
&N/A
&0.872
&0.904
&N/A
\\
GCN&SD
&0.472
&0.464
&N/A
&0.792
&0.728
&N/A
&0.84
&0.768
&N/A
\\
GCN&GC
&0.408
&0.368
&N/A
&0.728
&0.768
&N/A
&0.832
&0.872
&N/A
\\
SGC&Rand.
&0.616
&N/A&N/A
&0.824
&N/A&N/A
&0.872
&N/A&N/A
\\
SGC&SD
&0.696
&N/A&N/A
&0.8
&N/A&N/A
&0.816
&N/A&N/A
\\
SGC&GC
&0.6
&N/A&N/A
&0.824
&N/A&N/A
&0.912
&N/A&N/A
\\
\end{tabular}
\end{table}
\begin{table}[h!]
\caption{Results of influence attacks against each classifier with the Cora dataset, with attacker budgets of 10, 30, and 50 edge perturbations. Results are included for Nettack (Net), FGA, and IG-FGSM (IG). For each classifier, we train with random (Rand.), \textsc{StratDegree} (SD), and \textsc{GreedyCover} (GC). Each entry is a probability of attack success, thus higher is better for the attacker and lower is better for the defender. To yield the most robust classifier, the defender picks the classifier/training method combination that minimizes the worst-case attack probability. These entries are listed in \textbf{bold}. Entries representing the most robust case for random training are in \textit{italic}. Entries listed as N/A did not finish in the allotted time. The Jaccard-based classifier performs best, both overall and using random training. If we focus on classifiers that achieve the best performance in Figure~\ref{fig:classification}, (i.e., omitting Jaccard and SVD), the best performance is achieved by GCNs with the alternative training methods.}
\label{tab:cora_indirect}
\footnotesize
  \begin{tabular}{lc|ccc|ccc|ccc}
\hline
&&\multicolumn{3}{c|}{Budget 10}
&\multicolumn{3}{c|}{Budget 30}
&\multicolumn{3}{c}{Budget 50}\\
defense&training&Net&FGA&IG&Net&FGA&IG&Net&FGA&IG\\
\hline
Jaccard&Rand.
&\textit{0.296}
&0.24
&N/A
&\textit{0.472}
&0.432
&N/A
&\textit{0.592}
&0.504
&N/A
\\
Jaccard&SD
&0.264
&0.192
&N/A
&\textbf{0.352}
&0.312
&N/A
&0.4
&\textbf{0.424}
&N/A
\\
Jaccard&GC
&0.224
&\textbf{0.256}
&N/A
&0.36
&0.392
&N/A
&0.448
&0.456
&N/A
\\
RGCN&Rand.
&0.68
&0.6
&N/A
&0.912
&0.944
&N/A
&0.952
&0.976
&N/A
\\
RGCN&SD
&0.448
&0.408
&N/A
&0.848
&0.784
&N/A
&0.904
&0.888
&N/A
\\
RGCN&GC
&0.44
&0.296
&N/A
&0.832
&0.808
&N/A
&0.92
&0.888
&N/A
\\
Cheb&Rand.
&0.448
&0.448
&N/A
&0.784
&0.808
&N/A
&0.872
&0.896
&N/A
\\
Cheb&SD
&0.656
&0.48
&N/A
&0.976
&0.816
&N/A
&0.976
&0.912
&N/A
\\
Cheb&GC
&0.488
&0.448
&N/A
&0.864
&0.88
&N/A
&0.928
&0.912
&N/A
\\
SVD&Rand.
&0.224
&0.32
&N/A
&0.536
&0.72
&N/A
&0.76
&0.872
&N/A
\\
SVD&SD
&0.296
&0.216
&N/A
&0.568
&0.536
&N/A
&0.76
&0.728
&N/A
\\
SVD&GC
&0.264
&0.264
&N/A
&0.584
&0.528
&N/A
&0.84
&0.768
&N/A
\\
median&Rand.
&0.544
&0.424
&N/A
&0.872
&0.808
&N/A
&0.944
&0.912
&N/A
\\
median&SD
&0.4
&0.424
&N/A
&0.8
&0.784
&N/A
&0.912
&0.888
&N/A
\\
median&GC
&0.24
&0.312
&N/A
&0.784
&0.84
&N/A
&0.896
&0.912
&N/A
\\
GAT&Rand.
&0.544
&0.552
&N/A
&0.904
&0.88
&N/A
&0.96
&0.952
&N/A
\\
GAT&SD
&0.608
&0.48
&N/A
&0.88
&0.816
&N/A
&0.968
&0.872
&N/A
\\
GAT&GC
&0.48
&0.352
&N/A
&0.872
&0.768
&N/A
&0.952
&0.952
&N/A
\\
GCN&Rand.
&0.568
&0.448
&N/A
&0.896
&0.896
&N/A
&0.936
&0.952
&N/A
\\
GCN&SD
&0.456
&0.32
&N/A
&0.752
&0.696
&N/A
&0.856
&0.872
&N/A
\\
GCN&GC
&0.384
&0.32
&N/A
&0.816
&0.776
&N/A
&0.896
&0.896
&N/A
\\
SGC&Rand.
&0.568
&N/A&N/A
&0.824
&N/A&N/A
&0.888
&N/A&N/A
\\
SGC&SD
&0.632
&N/A&N/A
&0.84
&N/A&N/A
&0.856
&N/A&N/A
\\
SGC&GC
&0.584
&N/A&N/A
&0.824
&N/A&N/A
&0.88
&N/A&N/A
\\
\end{tabular}
\end{table}
\begin{table}[h!]
\caption{Results of influence attacks against each classifier with the PolBlogs dataset, with attacker budgets of 10, 30, and 50 edge perturbations. Results are included for Nettack (Net), FGA, and IG-FGSM (IG). For each classifier, we train with random (Rand.), \textsc{StratDegree} (SD), and \textsc{GreedyCover} (GC). Each entry is a probability of attack success, thus higher is better for the attacker and lower is better for the defender. To yield the most robust classifier, the defender picks the classifier/training method combination that minimizes the worst-case attack probability. These entries are listed in \textbf{bold}. Entries representing the most robust case for random training are in \textit{italic}. Entries listed as N/A did not finish in the allotted time. Best results overall and with random training are achieved with SVD, while RGCN performs equally well when using \textsc{StratDegree}.}
\label{tab:polblogs_indirect}
\footnotesize
  \begin{tabular}{cc|ccc|ccc|ccc}
\hline
&&\multicolumn{3}{c|}{Budget 10}
&\multicolumn{3}{c|}{Budget 30}
&\multicolumn{3}{c}{Budget 50}\\
defense&training&Net&FGA&IG&Net&FGA&IG&Net&FGA&IG\\
\hline
Jaccard&Rand.
&0.92
&0.872
&0.928
&0.992
&1.0
&1.0
&1.0
&1.0
&1.0
\\
Jaccard&SD
&0.928
&0.984
&0.936
&1.0
&1.0
&1.0
&1.0
&1.0
&1.0
\\
Jaccard&GC
&0.936
&0.952
&0.944
&0.992
&1.0
&1.0
&1.0
&1.0
&1.0
\\
RGCN&Rand.
&0.08
&0.128
&0.032
&0.224
&0.248
&0.056
&0.32
&0.304
&0.064
\\
RGCN&SD
&\textbf{0.048}
&0.024
&0.016
&\textbf{0.088}
&0.072
&0.056
&\textbf{0.112}
&0.096
&0.064
\\
RGCN&GC
&0.088
&0.096
&0.04
&0.192
&0.176
&0.048
&0.232
&0.272
&0.056
\\
Cheb&Rand.
&0.048
&0.184
&0.112
&0.184
&0.376
&0.176
&0.296
&0.456
&0.216
\\
Cheb&SD
&0.128
&0.128
&0.168
&0.256
&0.232
&0.288
&0.368
&0.32
&0.336
\\
Cheb&GC
&0.344
&0.08
&0.168
&0.456
&0.208
&0.328
&0.52
&0.288
&0.392
\\
SVD&Rand.
&0.016
&\textit{0.08}
&0.048
&0.04
&\textit{0.12}
&0.096
&0.112
&\textit{0.136}
&0.128
\\
SVD&SD
&0.064
&0.024
&0.064
&\textbf{0.088}
&0.024
&0.072
&\textbf{0.112}
&0.032
&0.096
\\
SVD&GC
&0.04
&0.04
&\textbf{0.048}
&\textbf{0.088}
&0.064
&0.08
&0.104
&0.112
&0.128
\\
GAT&Rand.
&0.224
&0.224
&0.184
&0.384
&0.408
&0.248
&0.448
&0.488
&0.304
\\
GAT&SD
&0.12
&0.056
&0.08
&0.184
&0.16
&0.152
&0.256
&0.208
&0.208
\\
GAT&GC
&0.112
&0.12
&0.064
&0.2
&0.256
&0.152
&0.288
&0.296
&0.168
\\
GCN&Rand.
&0.16
&0.208
&0.128
&0.288
&0.368
&0.2
&0.344
&0.424
&0.232
\\
GCN&SD
&0.104
&0.096
&0.176
&0.168
&0.208
&0.248
&0.208
&0.288
&0.312
\\
GCN&GC
&0.072
&0.032
&0.056
&0.152
&0.184
&0.104
&0.28
&0.296
&0.128
\\
SGC&Rand.
&N/A
&N/A&N/A
&N/A
&N/A&N/A
&N/A
&N/A&N/A
\\
SGC&SD
&N/A
&N/A&N/A
&N/A
&N/A&N/A
&N/A
&N/A&N/A
\\
SGC&GC
&0.056
&N/A&N/A
&0.12
&N/A&N/A
&0.168
&N/A&N/A
\\
\end{tabular}
\end{table}
\begin{table}[h!]
\caption{Results of influence attacks against each classifier with the PubMed dataset, with attacker budgets of 10, 30, and 50 edge perturbations. Results are included for Nettack (Net), FGA, and IG-FGSM (IG). For each classifier, we train with random (Rand.), \textsc{StratDegree} (SD), and \textsc{GreedyCover} (GC). Each entry is a probability of attack success, thus higher is better for the attacker and lower is better for the defender. To yield the most robust classifier, the defender picks the classifier/training method combination that minimizes the worst-case attack probability. These entries are listed in \textbf{bold}. Entries listed as N/A did not finish in the allotted time. Only results using Jaccard, GCN, and ChebNet were obtained in time. While \textsc{StratDegree} and \textsc{GreedyCover} improve performance with the Jaccard-based classifier, the best performance is achieved by a ChebNet classifier with random training. In our experiments, this classifier with the PubMed data typically has a much higher margin before the attack takes place.}
\label{tab:pubmed_indirect}
\footnotesize
  \begin{tabular}{lc|ccc|ccc|ccc}
\hline
&&\multicolumn{3}{c|}{Budget 10}
&\multicolumn{3}{c|}{Budget 30}
&\multicolumn{3}{c}{Budget 50}\\
defense&training&Net&FGA&IG&Net&FGA&IG&Net&FGA&IG\\
\hline
Jaccard&Rand.
&0.224
&N/A
&N/A
&0.576
&N/A
&N/A
&0.744
&N/A
&N/A
\\
Jaccard&SD
&0.12
&N/A
&N/A
&0.136
&N/A
&N/A
&0.16
&N/A
&N/A
\\
Jaccard&GC
&0.128
&N/A
&N/A
&0.192
&N/A
&N/A
&0.2
&N/A
&N/A
\\
GCN&Rand.
&0.456
&N/A
&N/A
&0.76
&N/A
&N/A
&0.888
&N/A
&N/A
\\
GCN&SD
&0.6
&N/A
&N/A
&0.936
&N/A
&N/A
&0.976
&N/A
&N/A
\\
GCN&GC
&0.544
&N/A
&N/A
&0.88
&N/A
&N/A
&0.952
&N/A
&N/A
\\
Cheb&Rand.
&\textbf{\textit{0.056}}
&N/A
&N/A
&\textbf{\textit{0.072}}
&N/A
&N/A
&\textbf{\textit{0.072}}
&N/A
&N/A
\\
Cheb&SD
&0.072
&N/A
&N/A
&0.128
&N/A
&N/A
&0.136
&N/A
&N/A
\\
Cheb&GC
&0.136
&N/A
&N/A
&0.16
&N/A
&N/A
&0.192
&N/A
&N/A
\\
\end{tabular}
\end{table}
\begin{table}[h!]
\caption{Results of direct attacks against each classifier with the CiteSeer dataset, with attacker budgets of 5, 10, and 20 edge perturbations. Results are included for Nettack (Net), FGA, and IG-FGSM (IG). For each classifier, we train with random (Rand.), \textsc{StratDegree} (SD), and \textsc{GreedyCover} (GC). Each entry is a probability of attack success, thus higher is better for the attacker and lower is better for the defender. To yield the most robust classifier, the defender picks the classifier/training method combination that minimizes the worst-case attack probability. These entries are listed in \textbf{bold}. Entries representing the most robust case for random training are in \textit{italic}. Entries listed as N/A did not finish in the allotted time. As with influence attacks, the Jaccard-based classifier performs best, though ChebNet also performs well for all training methods.}
\label{tab:citeseer_direct}
\footnotesize
  \begin{tabular}{lc|ccc|ccc|ccc}
\hline
&&\multicolumn{3}{c|}{Budget 5}
&\multicolumn{3}{c|}{Budget 10}
&\multicolumn{3}{c}{Budget 20}\\
defense&training&Net&FGA&IG&Net&FGA&IG&Net&FGA&IG\\
\hline
Jaccard&Rand.
&0.384
&0.256
&0.296
&0.592
&0.392
&0.368
&0.752
&0.464
&0.472
\\
Jaccard&SD
&0.432
&0.32
&0.256
&0.672
&0.416
&0.32
&0.808
&0.52
&0.432
\\
Jaccard&GC
&0.2
&0.184
&\textbf{0.224}
&0.264
&\textbf{0.336}
&\textbf{0.336}
&0.408
&0.52
&0.456
\\
RGCN&Rand.
&0.976
&0.848
&0.8
&0.992
&0.936
&0.936
&1.0
&0.968
&0.96
\\
RGCN&SD
&0.88
&0.568
&0.912
&0.992
&0.856
&0.976
&1.0
&0.944
&0.976
\\
RGCN&GC
&0.896
&0.712
&0.808
&1.0
&0.88
&0.936
&1.0
&0.952
&0.976
\\
Cheb&Rand.
&0.224
&\textit{0.28}
&0.264
&0.344
&0.352
&\textit{0.384}
&0.424
&0.44
&\textit{0.464}
\\
Cheb&SD
&0.264
&0.24
&0.304
&0.32
&0.336
&0.392
&0.4
&0.376
&\textbf{0.448}
\\
Cheb&GC
&0.288
&0.24
&0.256
&0.376
&0.312
&0.328
&0.472
&0.4
&0.4
\\
SVD&Rand.
&0.552
&0.392
&0.768
&0.76
&0.576
&0.936
&0.944
&0.856
&0.952
\\
SVD&SD
&0.84
&0.544
&0.936
&0.984
&0.696
&0.976
&1.0
&0.856
&0.976
\\
SVD&GC
&0.408
&0.312
&0.864
&0.688
&0.488
&0.952
&0.968
&0.792
&0.992
\\
median&Rand.
&0.808
&0.792
&0.792
&0.96
&0.936
&0.96
&0.984
&0.952
&0.984
\\
median&SD
&0.904
&0.84
&0.872
&0.992
&0.952
&0.952
&1.0
&0.96
&0.952
\\
median&GC
&0.856
&0.832
&0.848
&0.96
&0.952
&0.96
&0.992
&0.968
&0.976
\\
GAT&Rand.
&0.92
&0.864
&0.84
&0.984
&0.952
&0.936
&1.0
&0.96
&0.952
\\
GAT&SD
&0.944
&0.808
&0.952
&1.0
&0.952
&0.992
&1.0
&0.984
&1.0
\\
GAT&GC
&0.936
&0.808
&0.832
&1.0
&0.952
&0.92
&1.0
&0.984
&0.96
\\
GCN&Rand.
&0.944
&0.872
&N/A
&0.992
&0.952
&N/A
&1.0
&0.976
&N/A
\\
GCN&SD
&0.984
&0.832
&N/A
&1.0
&0.968
&N/A
&1.0
&0.992
&N/A
\\
GCN&GC
&0.912
&0.904
&0.936
&1.0
&0.976
&0.992
&1.0
&0.976
&0.992
\\
SGC&Rand.
&0.832
&N/A&N/A
&0.944
&N/A&N/A
&1.0
&N/A&N/A
\\
SGC&SD
&0.936
&N/A&N/A
&1.0
&N/A&N/A
&1.0
&N/A&N/A
\\
SGC&GC
&0.88
&N/A&N/A
&0.96
&N/A&N/A
&1.0
&N/A&N/A
\\
\end{tabular}
\end{table}
\begin{table}[h!]
\caption{Results of direct attacks against each classifier with the Cora dataset, with attacker budgets of 5, 10, and 20 edge perturbations. Results are included for Nettack (Net), FGA, and IG-FGSM (IG). For each classifier, we train with random (Rand.), \textsc{StratDegree} (SD), and \textsc{GreedyCover} (GC). Each entry is a probability of attack success, thus higher is better for the attacker and lower is better for the defender. To yield the most robust classifier, the defender picks the classifier/training method combination that minimizes the worst-case attack probability. These entries are listed in \textbf{bold}. Entries representing the most robust case for random training are in \textit{italic}. Entries listed as N/A did not finish in the allotted time. While random training with the SVD classifier works best at a low attack budget, Jaccard with \textsc{StratDegree} performs better against better-resourced attackers.}
\label{tab:cora_direct}
\footnotesize
  \begin{tabular}{lc|ccc|ccc|ccc}
\hline
&&\multicolumn{3}{c|}{Budget 5}
&\multicolumn{3}{c|}{Budget 10}
&\multicolumn{3}{c}{Budget 20}\\
defense&training&Net&FGA&IG&Net&FGA&IG&Net&FGA&IG\\
\hline
Jaccard&Rand.
&0.504
&0.328
&N/A
&\textit{0.712}
&0.48
&N/A
&\textit{0.952}
&0.68
&N/A
\\
Jaccard&SD
&0.448
&0.216
&N/A
&\textbf{0.656}
&0.36
&N/A
&\textbf{0.776}
&0.552
&N/A
\\
Jaccard&GC
&0.528
&0.296
&N/A
&0.76
&0.424
&N/A
&0.912
&0.616
&N/A
\\
RGCN&Rand.
&0.936
&0.896
&N/A
&0.984
&0.992
&N/A
&0.992
&0.992
&N/A
\\
RGCN&SD
&0.944
&0.832
&N/A
&1.0
&0.952
&N/A
&1.0
&0.96
&N/A
\\
RGCN&GC
&0.976
&0.84
&0.832
&1.0
&0.96
&0.976
&1.0
&0.96
&0.976
\\
Cheb&Rand.
&0.88
&0.728
&N/A
&0.976
&0.928
&N/A
&0.984
&0.96
&N/A
\\
Cheb&SD
&0.96
&0.752
&N/A
&1.0
&0.928
&N/A
&1.0
&0.936
&N/A
\\
Cheb&GC
&0.944
&0.816
&N/A
&0.992
&0.952
&N/A
&1.0
&0.96
&N/A
\\
SVD&Rand.
&\textbf{0.36}
&0.24
&N/A
&0.776
&0.592
&N/A
&0.992
&0.928
&N/A
\\
SVD&SD
&0.696
&0.288
&N/A
&0.936
&0.632
&N/A
&1.0
&0.92
&N/A
\\
SVD&GC
&0.432
&0.184
&N/A
&0.792
&0.448
&N/A
&1.0
&0.84
&N/A
\\
median&Rand.
&0.936
&0.768
&0.824
&0.992
&0.96
&0.968
&1.0
&0.976
&0.992
\\
median&SD
&0.968
&0.864
&0.864
&1.0
&0.952
&0.984
&1.0
&0.952
&0.984
\\
median&GC
&0.912
&0.824
&0.824
&1.0
&0.968
&0.976
&1.0
&0.968
&0.976
\\
GAT&Rand.
&0.944
&0.856
&N/A
&1.0
&0.96
&N/A
&1.0
&0.968
&N/A
\\
GAT&SD
&0.928
&0.736
&N/A
&1.0
&0.92
&N/A
&1.0
&0.968
&N/A
\\
GAT&GC
&0.92
&0.824
&N/A
&1.0
&0.952
&N/A
&1.0
&0.984
&N/A
\\
GCN&Rand.
&0.928
&0.888
&N/A
&0.992
&0.976
&N/A
&1.0
&0.976
&N/A
\\
GCN&SD
&0.928
&0.624
&0.904
&0.992
&0.944
&0.992
&1.0
&0.968
&0.992
\\
GCN&GC
&0.904
&0.832
&0.808
&0.992
&0.976
&0.984
&1.0
&0.984
&0.992
\\
SGC&Rand.
&0.896
&N/A&N/A
&1.0
&N/A&N/A
&1.0
&N/A&N/A
\\
SGC&SD
&0.944
&N/A&N/A
&1.0
&N/A&N/A
&1.0
&N/A&N/A
\\
SGC&GC
&0.904
&N/A&N/A
&1.0
&N/A&N/A
&1.0
&N/A&N/A
\\
\end{tabular}
\end{table}
\begin{table}[h!]
\caption{Results of direct attacks against each classifier with the PolBlogs dataset, with attacker budgets of 5, 10, and 20 edge perturbations. Results are included for Nettack (Net), FGA, and IG-FGSM (IG). For each classifier, we train with random (Rand.), \textsc{StratDegree} (SD), and \textsc{GreedyCover} (GC). Each entry is a probability of attack success, thus higher is better for the attacker and lower is better for the defender. To yield the most robust classifier, the defender picks the classifier/training method combination that minimizes the worst-case attack probability. These entries are listed in \textbf{bold}. Entries representing the most robust case for random training are in \textit{italic}. Entries listed as N/A did not finish in the allotted time. While random training with the SVD classifier works best at a low attack budget, Jaccard with \textsc{StratDegree} performs better against better-resourced attackers.}
\label{tab:polblogs_direct}
\footnotesize
  \begin{tabular}{lc|ccc|ccc|ccc}
\hline
&&\multicolumn{3}{c|}{Budget 5}
&\multicolumn{3}{c|}{Budget 10}
&\multicolumn{3}{c}{Budget 20}\\
defense&training&Net&FGA&IG&Net&FGA&IG&Net&FGA&IG\\
\hline
Jaccard&Rand.
&0.672
&0.616
&0.688
&0.896
&0.848
&0.856
&0.992
&0.968
&0.976
\\
Jaccard&SD
&0.752
&0.8
&0.808
&0.936
&0.928
&0.952
&1.0
&0.984
&1.0
\\
Jaccard&GC
&0.768
&0.76
&0.872
&0.904
&0.92
&0.976
&0.992
&1.0
&1.0
\\
RGCN&Rand.
&0.48
&0.464
&0.312
&0.632
&0.664
&0.504
&0.784
&0.872
&0.704
\\
RGCN&SD
&0.304
&0.28
&0.344
&0.4
&0.44
&0.52
&0.648
&0.568
&0.736
\\
RGCN&GC
&0.56
&0.488
&0.336
&0.72
&0.648
&0.456
&0.832
&0.832
&0.64
\\
Cheb&Rand.
&0.376
&0.352
&0.448
&0.536
&0.552
&0.544
&0.72
&0.68
&0.744
\\
Cheb&SD
&0.408
&0.376
&0.432
&0.624
&0.544
&0.584
&0.784
&0.688
&0.768
\\
Cheb&GC
&0.576
&0.424
&0.472
&0.68
&0.568
&0.592
&0.8
&0.728
&0.768
\\
SVD&Rand.
&0.184
&0.056
&\textit{0.336}
&0.368
&0.104
&0.552
&0.496
&0.192
&0.704
\\
SVD&SD
&0.288
&0.016
&0.368
&0.416
&0.024
&0.536
&0.496
&0.064
&0.808
\\
SVD&GC
&0.12
&0.04
&\textbf{0.32}
&0.36
&0.04
&0.512
&0.464
&0.048
&0.768
\\
GAT&Rand.
&0.456
&0.52
&0.376
&0.624
&0.792
&0.52
&0.824
&0.912
&0.672
\\
GAT&SD
&0.32
&0.24
&0.36
&0.464
&0.384
&0.504
&0.664
&0.648
&0.752
\\
GAT&GC
&0.552
&0.528
&0.336
&0.744
&0.84
&0.576
&0.856
&0.936
&0.84
\\
GCN&Rand.
&0.472
&0.624
&0.408
&0.68
&0.784
&0.52
&0.816
&0.92
&0.76
\\
GCN&SD
&0.424
&0.344
&0.408
&0.568
&0.528
&0.544
&0.76
&0.76
&0.672
\\
GCN&GC
&0.496
&0.552
&0.312
&0.72
&0.768
&0.512
&0.88
&0.912
&0.728
\\
SGC&Rand.
&0.36
&N/A&N/A
&\textbf{0.464}
&N/A&N/A
&\textbf{0.6}
&N/A&N/A
\\
SGC&SD
&N/A
&N/A&N/A
&N/A
&N/A&N/A
&N/A
&N/A&N/A
\\
SGC&GC
&0.528
&N/A&N/A
&0.736
&N/A&N/A
&0.856
&N/A&N/A
\\
\end{tabular}
\end{table}
\begin{table}[h!]
\caption{Results of direct attacks against each classifier with the PubMed dataset, with attacker budgets of 5, 10, and 20 edge perturbations. Results are included for Nettack (Net), FGA, and IG-FGSM (IG). For each classifier, we train with random (Rand.), \textsc{StratDegree} (SD), and \textsc{GreedyCover} (GC). Each entry is a probability of attack success, thus higher is better for the attacker and lower is better for the defender. To yield the most robust classifier, the defender picks the classifier/training method combination that minimizes the worst-case attack probability. These entries are listed in \textbf{bold}. Entries representing the most robust case for random training are in \textit{italic}. Entries listed as N/A did not finish in the allotted time. While the a\textsc{StratDegree} \textsc{GreedyCover} work well in conjunction with the Jaccard classifier, a disparity in the classification margin hinders their performance in the best case, using a ChebNet classifier.}
\label{tab:pubmed_direct}
\footnotesize
  \begin{tabular}{lc|ccc|ccc|ccc}
\hline
&&\multicolumn{3}{c|}{Budget 5}
&\multicolumn{3}{c|}{Budget 10}
&\multicolumn{3}{c}{Budget 20}\\
defense&training&Net&FGA&IG&Net&FGA&IG&Net&FGA&IG\\
\hline
Jaccard&random
&0.784
&N/A
&N/A
&0.952
&N/A
&N/A
&0.992
&N/A
&N/A
\\
Jaccard&degree
&0.208
&N/A
&N/A
&0.248
&N/A
&N/A
&0.328
&N/A
&N/A
\\
Jaccard&cover
&0.208
&N/A
&N/A
&0.328
&N/A
&N/A
&0.456
&N/A
&N/A
\\
GCN&random
&0.92
&N/A
&N/A
&1.0
&N/A
&N/A
&1.0
&N/A
&N/A
\\
GCN&degree
&0.952
&N/A
&N/A
&1.0
&N/A
&N/A
&1.0
&N/A
&N/A
\\
GCN&cover
&0.936
&N/A
&N/A
&0.984
&N/A
&N/A
&1.0
&N/A
&N/A
\\
Cheb&random
&\textbf{0.056}
&N/A
&N/A
&\textbf{0.088}
&N/A
&N/A
&\textbf{0.088}
&N/A
&N/A
\\
Cheb&degree
&0.072
&N/A
&N/A
&\textbf{0.088}
&N/A
&N/A
&0.104
&N/A
&N/A
\\
Cheb&cover
&0.088
&N/A
&N/A
&0.112
&N/A
&N/A
&0.152
&N/A
&N/A
\\
GAT&random
&0.792
&N/A
&N/A
&0.896
&N/A
&N/A
&0.992
&N/A
&N/A
\\
GAT&degree
&0.936
&N/A
&N/A
&0.992
&N/A
&N/A
&1.0
&N/A
&N/A
\\
GAT&cover
&0.92
&N/A
&N/A
&0.984
&N/A
&N/A
&0.992
&N/A
&N/A
\\
\end{tabular}
\end{table}

\end{document}